\documentclass[a4paper,11pt]{article}
\pdfoutput=1 
\usepackage{jheppub} 
\usepackage{amssymb}
\usepackage{epsfig}
\usepackage{graphicx}%
\usepackage{amsmath}%

\title{\boldmath Quantitative Study \\ of Different Forms of Geometrical Scaling \\in Deep Inelastic Scattering at HERA}
\author{Michal Praszalowicz}
\author{and Tomasz Stebel}
\affiliation{M. Smoluchowski Institute of Physics, Jagiellonian University,\\
Reymonta 4, 30-059 Krak{\'o}w, Poland}

\emailAdd{michal@if.uj.edu.pl}
\emailAdd{tomasz.stebel@uj.edu.pl}

\abstract{We use recently proposed {\em method of ratios} to assess the quality of geometrical
scaling in deep inelastic scattering for different forms of the saturation scale.  We consider original form of
geometrical scaling  (motivated by the Balitski-Kovchegov (BK) equation with {\em fixed coupling})
studied in more detail in our previous paper, and four new hypotheses: phenomenologically
motivated case with $Q^2$ dependent exponent $\lambda$ that governs small $x$ dependence of the saturation scale, two versions of  scaling ({\em running coupling 1 and 2}) that follow from the BK equation with running coupling, and {\em diffusive scaling} suggested by the QCD evolution equation beyond mean field approximation. It turns out that more sophisticated scenarios: {\em running coupling} scaling and {\em diffusive scaling} are disfavored by the combined HERA data  on $e^+p$ deep inelastic structure function $F_2$.}

\begin{document}
\maketitle
\flushbottom

\section{Introduction}

\label{intro}

Geometrical scaling (GS) has been introduced in Ref.~\cite{Stasto:2000er} in
the context of low $x$ Deep Inelastic Scattering (DIS). It has been
conjectured that $\gamma^{\ast}p$ cross-section $\sigma_{\gamma^{\ast}%
p}(x,Q^{2})=4\pi^{2}\alpha_{\mathrm{em}}F_{2}(x,Q^{2})/Q^{2}$ which in
principle depends on two independent kinematical variables $Q^{2}$ and $W$
(\emph{i.e.} $\gamma^{\ast}p$ scattering energy), depends only on a specific
combination of them, namely upon%
\begin{equation}
\tau=\frac{Q^{2}}{Q_{\text{s}}^{2}(x)} \label{tau}%
\end{equation}
called scaling variable. Bjorken $x$ variable is defined as
\begin{equation}
x=\frac{Q^{2}}{Q^{2}+W^{2}-M_{\text{p}}^{2}} \label{xdef}%
\end{equation}
and $M_{p}$ denotes the proton mass. In Ref.~\cite{Stasto:2000er}, following
Golec-Biernat--W{\"u}sthoff  (GBW) model \cite{GolecBiernat:1998js},
function $Q_{\text{s}}(x)$ -- called
saturation scale --  was taken in the following form
\begin{equation}
Q_{\text{s}}^{2}(x)=Q_{0}^{2}\left(  \frac{x}{x_{0}}\right)  ^{-\lambda}.
\label{Qsat}%
\end{equation}
Here $Q_{0}$ and $x_{0}$ are free parameters which can be extracted from the
data within some specific model of DIS, and exponent $\lambda$ is a dynamical
quantity of the order of $\lambda\sim0.3$. In the GBW model $Q_{0}=1~$GeV$/c$ 
and $x_{0}=3 \times10^{-4}.$

In our previous paper \cite{Praszalowicz:2012zh} (see also \cite{Stebel:2012ky}) we have
proposed a simple \emph{method of ratios} to assess in the model independent way the quality and 
the range of applicability of GS 
for the saturation scale defined in  Eq.~(\ref{Qsat}). Here we follow the same steps
to test four different forms of the saturation scale that have been proposed in the
literature.

Geometrical scaling is theoretically motivated by the gluon saturation
phenomenon (for review see Refs.~\cite{Mueller:2001fv,McLerran:2010ub})
in which low $x$ gluons of given transverse size $\sim1/Q^{2}$
start to overlap and their number is no longer growing once $Q^{2}$ is
decreased. This phenomenon -- called gluon saturation --  appears 
formally due to the nonlinearities of parton evolution 
at small $x$ given
by so called JIMWLK hierarchy equations \cite{jimwlk} which in the large
$N_{\rm c}$ limit reduce to the Balitsky-Kovchegov equation \cite{BK}.
These equations admit traveling wave solutions which explicitly exhibit
GS \cite{Munier:2003vc}. An effective theory describing small $x$ regime 
is Color Glass Condensate \cite{sat1,MLV}.

Gluon saturation takes place for Bjorken
$x$ much smaller than 1. Yet in Ref.~\cite{Praszalowicz:2012zh} we have shown
that GS with saturation scale defined by Eq.~(\ref{Qsat}) works very well up
to much higher values of $x$, namely up to $x \sim 0.1$. In this region GS cannot be
attributed to the saturation physics alone. Indeed, it is known that GS scaling extends 
well above the saturation scale both in the Dokshitzer-Gribov-Lipatov--Altarelli-Parisi
 \cite{DGLAP} (DGLAP) \cite{Kwiecinski:2002ep} and Balitsky-Lipatov-Fadin-Kuraev 
 \cite{BFKL} (BFKL) \cite{Iancu:2002tr} evolution schemes once the boundary conditions 
satisfy GS to start with. It has been also shown that in DGLAP scheme GS builds up during
evolution for generic boundary conditions \cite{Caola:2008xr}. Therefore in the
kinematical region far from the saturation regime where, however, no other scales exist 
({\em e.g.} for nearly massless particles) it is still  the saturation scale which governs
the behavior of the $\gamma^{\ast}p$ cross-section.

The form of saturation scale given by Eq.(\ref{Qsat}) is dictated
by the asymptotic behavior \cite{Munier:2003vc}
of the Balitsky-Kovchegov (BK) equation \cite{BK}, 
which is essentially the BFKL equation \cite{BFKL} 
supplied with a nonlinear damping term. It has been first used in the papers by 
K. Golec-Biernat and M. W\"{u}sthoff  \cite{GolecBiernat:1998js} 
where the saturation model of inclusive and
diffractive DIS has been formulated and tested phenomenologically.

Since the original discovery of GS in 2001 there have been many theoretical
attempts to find a "better" scaling variable which is both theoretically
justified and phenomenologically acceptable. An immediate generalization of
the saturation model of Refs.~\cite{GolecBiernat:1998js} has been done in 
Ref.~\cite{Bartels:2002cj} where DGLAP \cite{DGLAP}
evolution in $Q^{2}$ has
been included. Although the exact formulation of DGLAP improved saturation
model requires numerical solution of DGLAP equations, one can take
this into account \emph{phenomenologically} by allowing for an effective  $Q^{2}$ 
dependence of
the exponent $\lambda=\lambda_{\rm phn}(Q^{2})$ which is indeed seen experimentally
in the low $x$ behavior of $F_{2}$ structure function (see {\em e.g.} 
Refs.~\cite{Bartels:2002cj,Kowalski:2010ue} and Fig.~\ref{lameff}). 
This piece of data can
be relatively well described by the linear dependence of $\lambda_{\rm phn}(Q^{2})$
on $\log Q^{2}$ leading to the scaling variable of the following form%
\begin{equation}
\tau_{\text{phn}}=Q^{2}x^{\lambda_{0}+\beta\log Q^{2}/Q_{\beta}^{2}} \label{eff}%
\end{equation}

In another approach to DIS at low $x$ one considers modifications of BK
equation through an inclusion of the running coupling constant effects.
Depending on the approximations used two different forms of scaling variable
have been discussed in the literature \cite{Munier:2003vc}:%
\begin{equation}
\tau_{\text{rc1}}= Q^{2} e^{{-\mu\sqrt{\log\left(  1/x\right)  }}} \label{rc1}%
\end{equation}
and \cite{Beuf:2008mb}%
\begin{equation}
\tau_{\text{rc2}}=Q^{2}x^{\nu/\log (Q^{2}/Q_{\nu}^{2})} \label{rc2}%
\end{equation}
where subscripts "rc" refer to "running coupling". Note that from
phenomenological point of view (\ref{rc2}) is in fact a variation of
(\ref{eff}) where a different form of $Q^{2}$ dependence has been used.
Finally, generalization of the BK equation beyond a mean-field approximation
leads to so called \emph{diffusive scaling} \cite{DiffScal} characterized by yet another
scaling variable:%
\begin{equation}
\tau_{\text{ds}}=\left(  Q^{2}\right)  ^{1/\sqrt{\log(1/x)}} e^{-\kappa
\sqrt{\log\left(  1/x\right)  }} . \label{ds}%
\end{equation}

These different forms of scaling variable (except (\ref{eff})) have been
tested in a series of papers \cite{Gelis:2006bs,Beuf:2008bb,Royon:2010tz}
where the so called Quality Factor (QF) has been defined and used as a tool to
assess the quality of geometrical scaling.  In the following 
we shall use the method developed in Refs.~\cite{Praszalowicz:2012zh,Stebel:2012ky} 
 to test hypothesis of
GS in scaling variables (\ref{eff})--(\ref{ds}) and to study the region of its
applicability using combined analysis of $e^+p$ HERA data \cite{HERAcombined}.
We shall also compare our results with earlier findings of
Refs.\cite{Gelis:2006bs,Beuf:2008bb,Royon:2010tz}. 

Our results can be
summarized as follows: more sophisticated scenarios  {\em i.e. running coupling} scaling and {\em diffusive scaling} are disfavored by the combined HERA data  on $e^+p$ deep inelastic structure function $F_2$.
In contrast, phenomenologically motivated case with $Q^2$ dependent exponent $\lambda$  and the originally proposed form of the saturation scale 
\cite{Stasto:2000er} with fixed $\lambda$ exhibit high quality geometrical scaling over the large
region of Bjorken $x$ up to 0.1. The fact that GS is valid up to much larger Bjorken $x$'s than
originally anticipated has been already used in an analysis of  GS in the multiplicity $p_{\rm T}$ spectra
in $pp$ collisions \cite{Praszalowicz:2013uu}.

In Sect.~\ref{method} we briefly recapitulate the \emph{method of ratios} of 
Ref.~\cite{Praszalowicz:2012zh}
and define the criteria for GS
to hold. In Sect.~\ref{res} we present results for 4 different scaling variables
introduced in Eqs.(\ref{eff})--(\ref{ds}). Finally in Sect.~\ref{sumcon} we compare these
results with our previous paper \cite{Praszalowicz:2012zh} and with the results of 
Refs.~\cite{Gelis:2006bs,Beuf:2008bb,Royon:2010tz}.

\section{Method of ratios}
\label{method}

Throughout this paper we shall use model-independent method used in 
Refs.~\cite{Stebel:2012ky,Praszalowicz:2012zh} which was developed
in Refs.\cite{GSinpp} to test GS in multiplicity distributions at the LHC. 
Geometrical scaling hypothesis means that%
\begin{equation}
\sigma_{\gamma^{\ast}p}(x_{i},Q^{2})=\frac{1}{Q_{0}^{2}}F(\tau) \label{GSdef}%
\end{equation}
where for simplicity we define $\sigma_{\gamma^{\ast}p}$ as
\begin{equation}
\sigma_{\gamma^{\ast}p}(x_{i}, Q^{2})=\frac{F_{2}(x_{i}, Q^{2})}{Q^{2}}.
\label{gampi}%
\end{equation}
Function $F$ in Eq.~(\ref{GSdef}) is a universal dimensionless function of
$\tau$. In view of Eq.~(\ref{GSdef}) cross-sections $\sigma_{\gamma^{\ast}%
p}(x_{i},Q^{2})$ for different $x_{i}$'s, evaluated not in terms of $Q^{2}$
but in terms of $\tau$, should fall on one universal curve. This means in turn
that if we calculate ratio of cross-sections for different Bjorken $x_{i}$'s,
each expressed in terms of $\tau$, we should get unity independently of $\tau
$. This allows to determine parameter governing $x$ dependence of $\tau$ by
minimizing deviations of these ratios from unity. Generically we denote this
parameter as $\alpha$, although for each scaling variable (\ref{eff}) --
(\ref{ds}) it has a different meaning: $\alpha=\beta,\,\mu,\,\nu$ and
$\kappa$ for $Q^{2}$-dependent, running coupling (1 and 2) and diffusive
scaling hypotheses, respectively.

Following \cite{Stebel:2012ky,Praszalowicz:2012zh} we apply here the following procedure. 
First we
choose some $x_{\mathrm{ref}}$ and consider all Bjorken $x_{i}$'s smaller than
$x_{\mathrm{ref}}$ that have at least two overlapping points in $Q^{2}$ (or
more precisely in scaling variable $\tau$). Next we form the ratios
\begin{equation}
R_{x_{i},x_{\text{ref}}}(\alpha;\tau_{k})=\frac{\sigma_{\gamma^{\ast}p}%
(x_{i},\tau(x_{i},Q_{k}^{2};\alpha))}{\sigma_{\gamma^{\ast}p}(x_{\text{ref}%
},\tau(x_{\text{ref}},Q_{k,\text{ref}}^{2};\alpha))} \label{Rxdef}%
\end{equation}
with
\begin{equation}
\tau_{k}=\tau(x_{i},Q_{k}^{2};\alpha)=\tau(x_{\text{ref}},Q_{k,\text{ref}%
}^{2};\alpha). \label{interpol}%
\end{equation}
By tuning $\alpha$ one can make $R_{x_{i},x_{\text{ref}}}(\alpha;\tau
_{k}) = 1 \pm \delta$ for all $\tau_{k}$ with accuracy of $\delta$ for which
following Ref.~\cite{Praszalowicz:2012zh} we take 3\%.

For $\alpha\neq0$ points of the same $Q^{2}$ but
different $x$'s correspond generally to different $\tau$'s. Therefore one has
to interpolate the reference cross-section $\sigma_{\gamma^{\ast}%
p}(x_{\text{ref}},\tau(x_{\text{ref}},Q^{2};\alpha))$ to $Q_{k,\text{ref}%
}^{2}$ such that $\tau(x_{\text{ref}},Q_{k,\text{ref}}^{2};\alpha)=\tau_{k}$
as indicated in Eq.~(\ref{interpol}). This procedure is described in detail in 
Refs.~\cite{Praszalowicz:2012zh,Stebel:2012ky}.

In order to find optimal value of parameter $\alpha$ that minimizes
deviations of ratios (\ref{Rxdef}) from unity we form the chi-square measure%
\begin{equation}
\chi_{x_{i},x_{\text{ref}}}^{2}(\alpha)=\frac{1}{N_{x_{i},x_{\text{ref}}}%
-1}{\displaystyle\sum\limits_{k\in x_{i}}}\frac{\left(  R_{x_{i}%
,x_{\text{ref}}}(\alpha;\tau_{k})-1\right)  ^{2}}{\Delta R_{x_{i}%
,x_{\text{ref}}}(\alpha;\tau_{k})^{2}} \label{chix1}%
\end{equation}
where the sum over $k$ extends over all points of given $x_{i}$ that have
overlap with $x_{\text{ref}}$ and ${N_{x_{i},x_{\text{ref}}}}$ is a number of
such points.

Finally, the errors entering formula (\ref{chix1}) are calculated using%
\begin{align}
& \Delta R_{x_{i},x_{\text{ref}}}(\alpha;\tau_{k})^{2} =   \label{err1} \\
& \left( \left( 
\frac{\Delta\sigma_{\gamma^{\ast}p}(x_{i},\tau(x_{i},Q_{k}^{2}))}{
\sigma_{\gamma^{\ast}p}(x_{i},\tau(x_{i},Q_{k}^{2}))}\right)^{2} 
+\left(  \frac{\Delta\sigma_{\gamma^{\ast}p}(x_{\text{ref}},\tau(x_{\text{ref}},Q_{k,\text{ref}}%
^{2}))}%
{\sigma_{\gamma^{\ast}p}(x_{\text{ref}},\tau(x_{\text{ref}},Q_{k,\text{ref}}^2))}%
\right)  ^{2}\right) R_{x_{i},x_{\text{ref}}}(\alpha; \tau_k)^2 
+ \delta^2 & \nonumber %
\end{align}
where $\Delta\sigma_{\gamma^{\ast}p}(\tau(x,Q^{2}))$ are experimental errors
(or interpolated experimental errors) of $\gamma^{\ast}p$ cross-sections
(\ref{gampi}). For more detailed discussion of errors see Ref.~\cite{Praszalowicz:2012zh}.

In this way, for each pair of available Bjorken variables $(x_{i}%
,x_{\mathrm{ref}})$, we compute the \emph{best} value of parameter
$\alpha$, denoted in the following by a subscript 
min:\footnote{Because it minimizes $\chi^2$.}
$\alpha
_{\mathrm{min}}(x_{i},x_{\mathrm{ref}})$ and the corresponding $\chi^{2}$. For
GS to hold we should find a region in $(x_{i},x_{\mathrm{ref}})$ half-plane
(note that by construction $x_{i} < x_{\mathrm{ref}}$) where $\alpha
_{\mathrm{min}}(x_{i},x_{\mathrm{ref}})$ is a constant independent of $x_{i}$
and $x_{\mathrm{ref}}$, and the corresponding $\chi_{x_{i},x_{\mathrm{ref}}%
}^{2}$ is small.

We shall also look for possible violations of GS in a more quantitative way.
In order to eliminate the dependence of $\alpha_{\text{min}}%
(x,x_{\mathrm{ref}})$ on the value of $x$, we introduce averages over $x$
(denoted in the following by $\left\langle \ldots\right\rangle $) minimizing
the following chi-square function:
\begin{equation}
\tilde{\chi}_{x_{\text{ref}}}^{2}(\left\langle \alpha\right\rangle )=\frac
{1}{N_{x_{\mathrm{ref}}}-1}{\displaystyle\sum\limits_{x<x_{\text{ref}}}}%
\frac{\left(  \alpha_{\text{min}}(x,x_{\text{ref}})-\left\langle
\alpha\right\rangle \right)  ^{2}}{\Delta\alpha_{\text{min}}(x,x_{\text{ref}%
})^{2}} \label{chitildex}%
\end{equation}
which gives the {\em best} value of $\langle\alpha\rangle$ denoted as
$\langle\alpha_{\mathrm{min}}(x_{\text{ref}})\rangle$. The sum in
(\ref{chitildex}) extends over all $x$'s such that $\alpha_{\text{min}%
}(x,x_{\text{ref}})$ exists and $N_{x_{\mathrm{ref}}}$ is the number of terms
in (\ref{chitildex}).

Since GS is expected to work for small $x$'s, the "average" value of scaling
parameter $\langle\alpha_{\mathrm{min}}(x_{\text{ref}})\rangle$ supplies an
information, up to what value of $x_{\mathrm{ref}}$ GS is still working. For
small $x_{\mathrm{ref}}$ we expect $\langle\alpha_{\mathrm{min}%
}(x_{\text{ref}})\rangle$ to be constant, whereas for larger values we expect
to see some dependence of $\langle\alpha_{\mathrm{min}}(x_{\text{ref}%
})\rangle$ on $x_{\mathrm{ref}}$. A word of warning is here in order. Even if
$\langle\alpha_{\mathrm{min}}(x_{\text{ref}})\rangle$ is a constant we have
to look at the corresponding value of $\chi^{2}$: too large $\chi^{2}$
obviously indicates violation of GS.

To quantify further the hypothesis of geometrical scaling we form yet another
chi-square function%
\begin{equation}
\overline{\chi}_{x_{\text{cut}}}^{2}(\left\langle \left\langle \alpha
\right\rangle \right\rangle )=\frac{1}{N_{x_{\text{cut}}}-1}{\displaystyle\sum
\limits_{x_{\text{ref}}\leq x_{\text{cut}}}}\,{\displaystyle\sum
\limits_{x<x_{\text{ref}}}}\frac{\left(  \alpha_{\text{min}}(x,x_{\text{ref}%
})-\left\langle \left\langle \alpha\right\rangle \right\rangle \right)  ^{2}%
}{\Delta\alpha_{\text{min}}(x,x_{\text{ref}})^{2}} \label{titilchi2}%
\end{equation}
which we minimize to obtain $\left\langle \left\langle \alpha_{\text{min}%
}(x_{\text{cut}})\right\rangle \right\rangle $.

Equation (\ref{titilchi2}) allows us to see how well one can fit $\left\langle
\alpha_{\text{min}}(x_{\text{ref}})\right\rangle $ with a constant $\alpha$
up to $x_{\text{ref}}=x_{\text{cut}}$. Were there any strong violations of GS
above some $x_{0}$, one should see a rise of $\left\langle \left\langle
\alpha_{\text{min}}(x_{\text{cut}})\right\rangle \right\rangle $ once
$x_{\text{cut}}$ becomes larger than $x_{0}$.

\section{Results}
\label{res}

Let us now come back to the discussion of different scaling variables defined
in Eqs.~(\ref{eff})~--~(\ref{ds}). All of them depend on one variational
parameter, which we constrain analyzing ratios (\ref{Rxdef}) for
combined HERA $e^{+} p$ DIS data \cite{HERAcombined}.

\begin{figure}[ptb]
\centering
\includegraphics[width=10cm,angle=0]{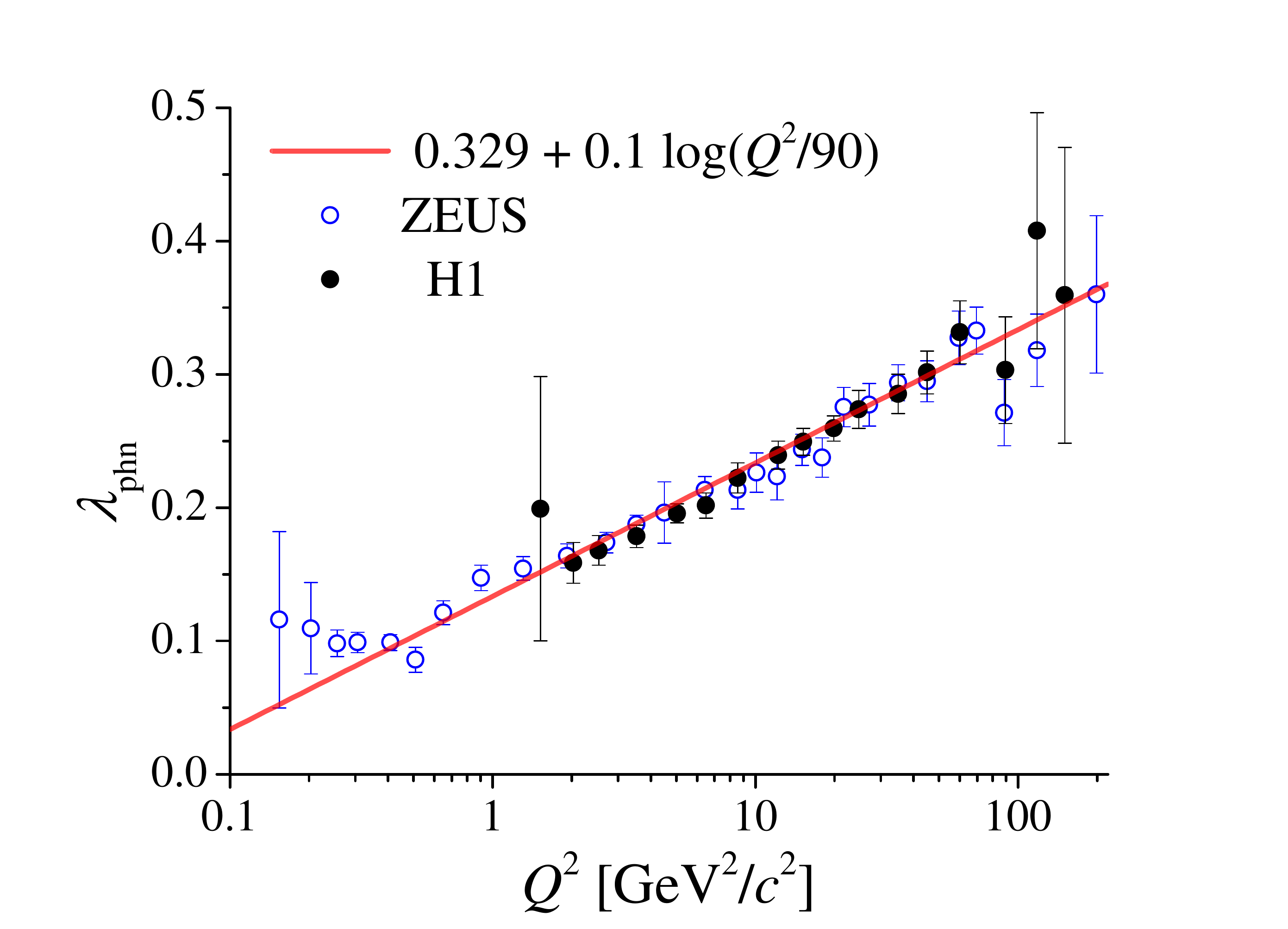} \caption{Effective exponent
$\lambda_{\rm phn}$ from $F_{2}$ at low $x$ (\ref{lowxF2}) from HERA 
and the linear fit of Eq.~(\ref{logfit}).  Data points as in Ref.~\cite{Bartels:2002cj},
see also \cite{Kowalski:2010ue}.}%
\label{lameff}%
\end{figure}

In the case of $Q^{2}$-dependent exponent $\lambda_{\rm phn}$ (\ref{eff}), however,
there are in fact two parameters, one of them ($\lambda_{0}$) being fixed
using our previous analysis of Ref.~\cite{Praszalowicz:2012zh}
 where we have shown that GS scaling
works very well with constant $\lambda=\lambda_0$: 
\begin{equation}
\lambda_{0}=0.329 \pm 0.002. \label{lambda0}
\end{equation} 

\begin{figure}[h!]
\centering
\hspace{0.3cm}
\includegraphics[width=7.5cm,angle=0]{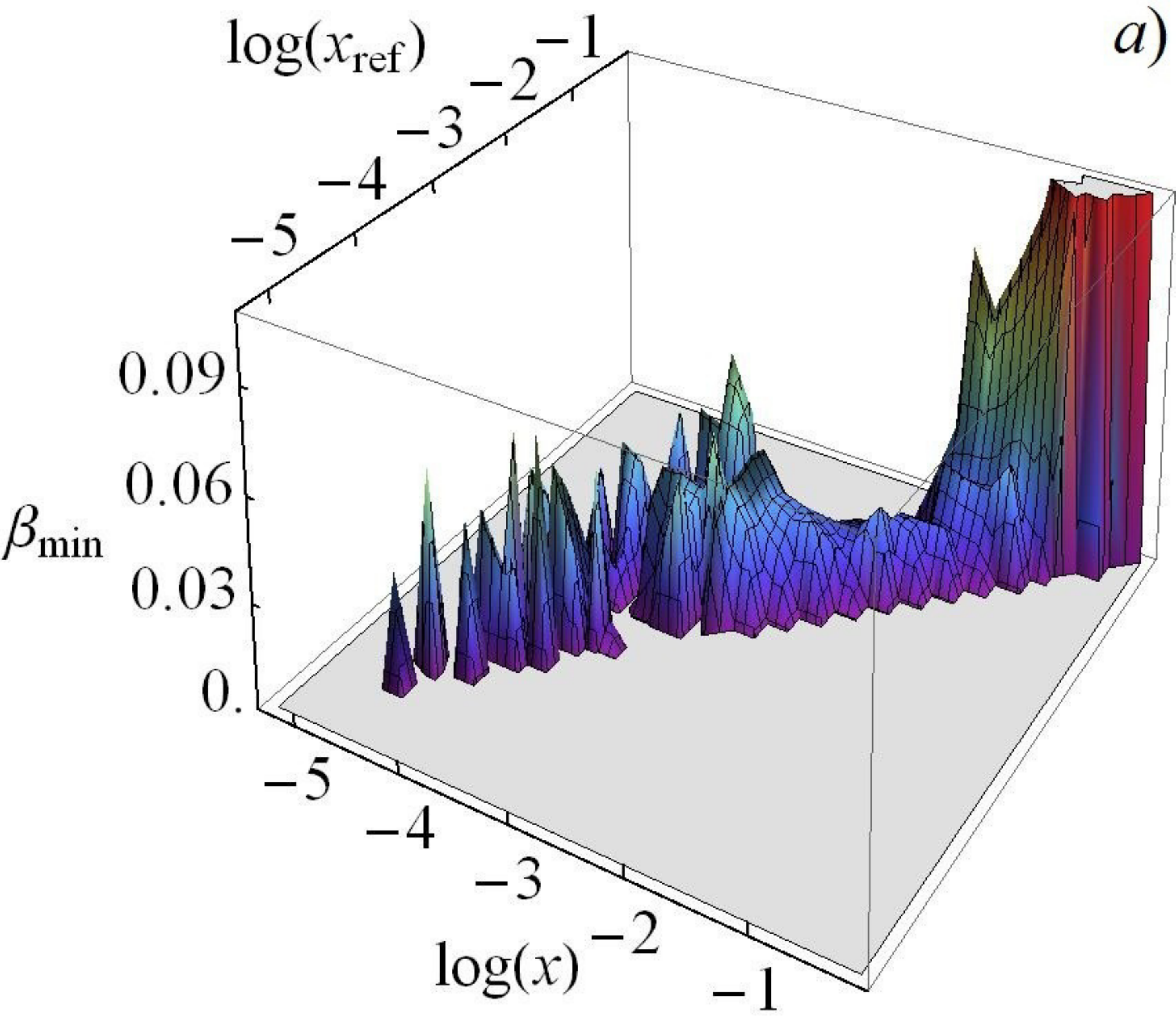}~~
\includegraphics[width=7cm,angle=0]{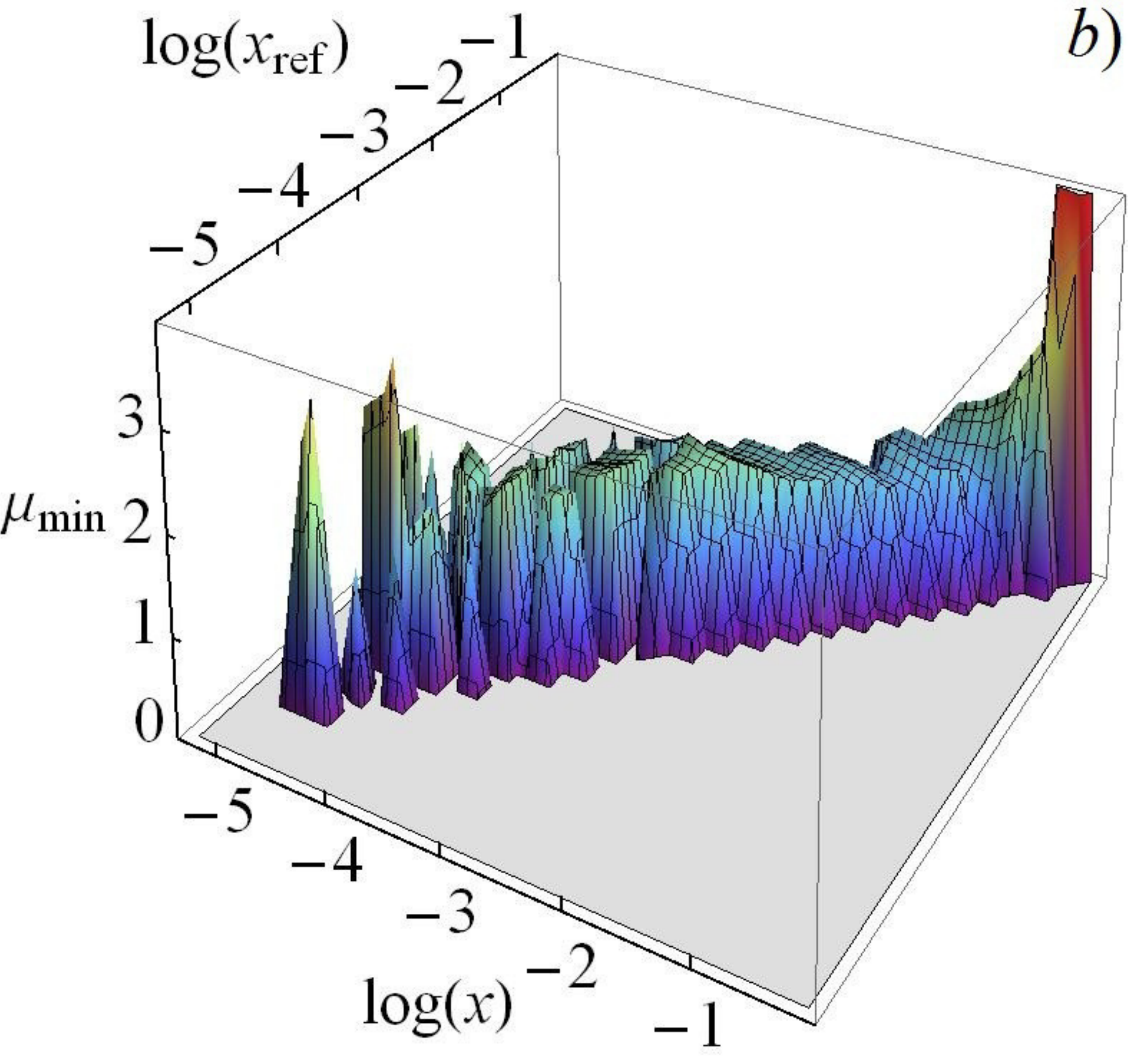} \\%
\vspace{0.5cm}
\includegraphics[width=7cm,angle=0]{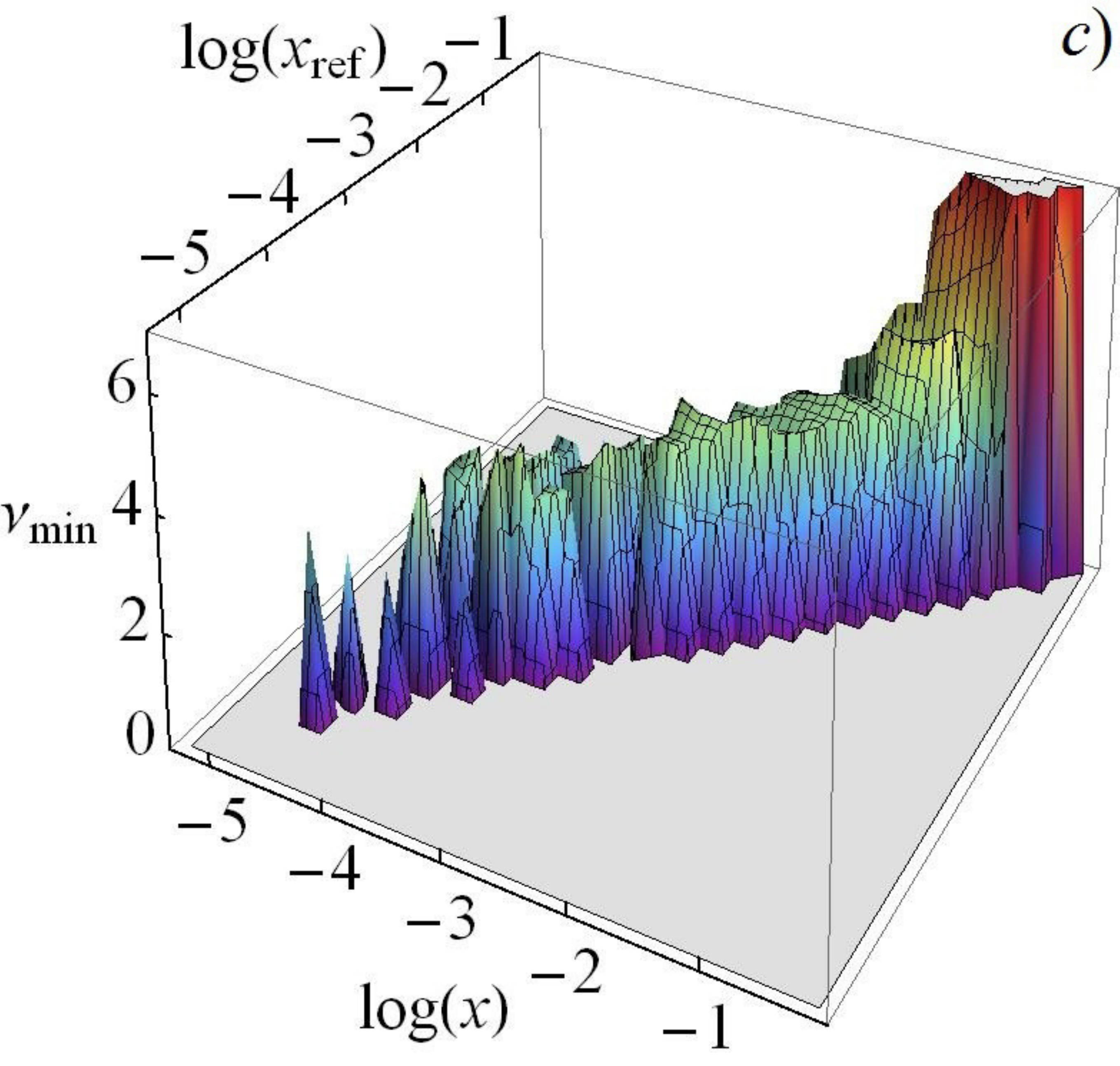}~~
\includegraphics[width=7cm,angle=0]{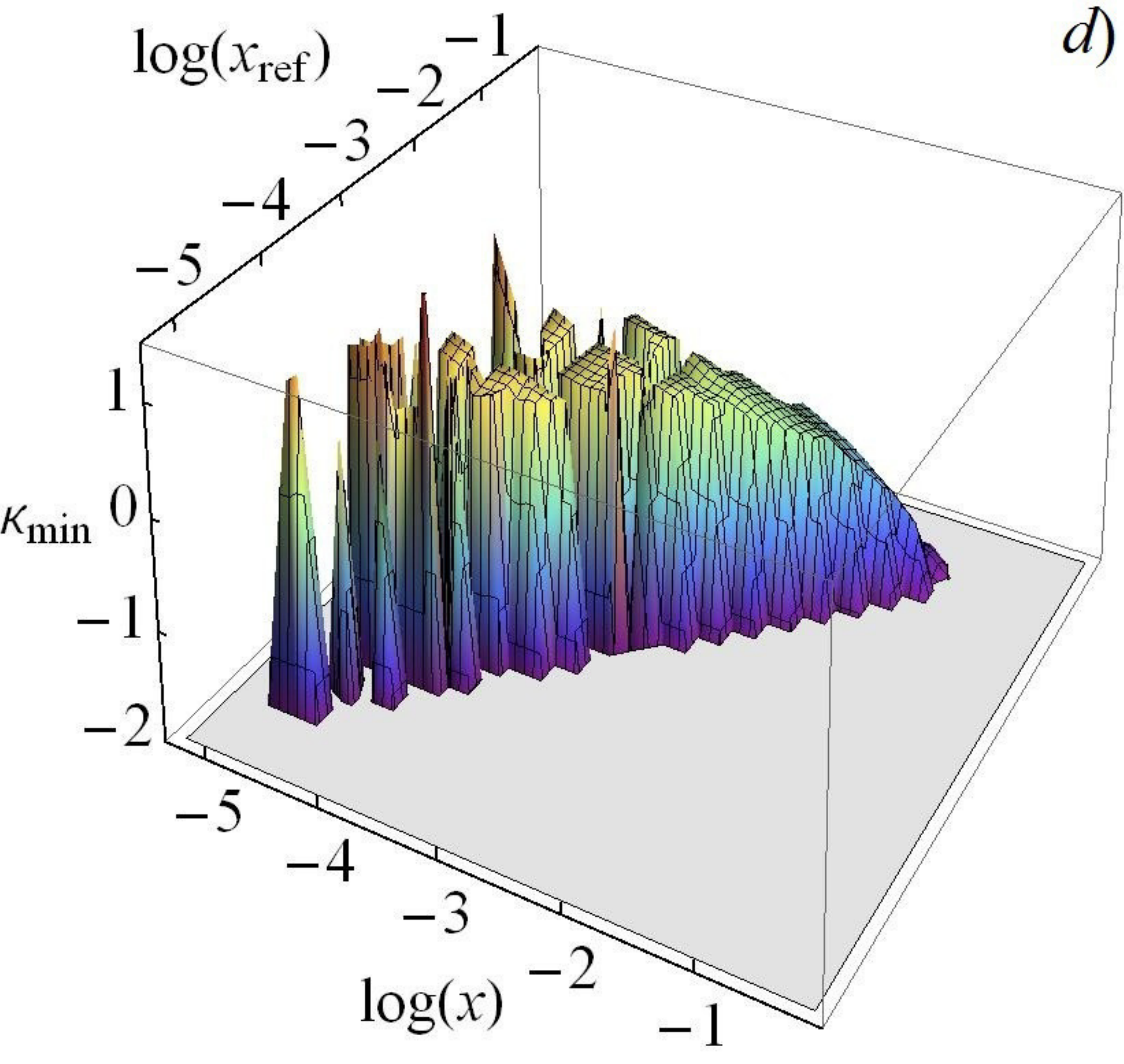}\\
\caption{Three dimensional plots of a) $\beta_{\rm min}(x,x_{\rm ref})$,
b) $\mu_{\rm min}(x,x_{\rm ref})$, c) $\nu_{\rm min}(x,x_{\rm ref})$
and d) $\kappa_{\rm min}(x,x_{\rm ref})$ obtained by minimizing $\chi^2$
function of Eq.~(\ref{chix1}).}%
\label{3dplots}%
\end{figure}

On the other hand
looking at low $x$ behavior of the $F_{2}$ structure function it has been
shown that \cite{Bartels:2002cj,Kowalski:2010ue}:
\begin{equation}
F_{2}(x,Q^{2})\sim x^{- \lambda_{\rm phn}(Q^{2})}
\label{lowxF2}%
\end{equation}
where $\lambda_{\rm phn}(Q^{2})$ can be well parametrized as
\begin{equation}
\lambda_{\rm phn}(Q^{2})=0.329+0.1\log(Q^{2}/90) \label{logfit}%
\end{equation}
(for $Q^{2}$ in (GeV/$c)^{2}$) as depicted in Fig.~\ref{lameff}.
Taking therefore scaling variable in the form of (\ref{eff}) with
$\lambda_{0}=0.329$ we test in fact consistency of the slopes $\beta$ as
extracted from Fig.~\ref{lameff} and by the procedure described in Sect.~\ref{method}.
Note that this is therefore a kind of perturbative two parameter fit, and as
such it has a different status than the remaining Ans\"{a}tze for the scaling
variable (\ref{rc1}) -- (\ref{ds}). Similar remarks apply to the running
coupling rc2 case (\ref{rc2}), where the scale of the logarithm $Q_{\nu}^2$ has been fixed at $0.04$
(following {\em e.g.} Ref.~\cite{Beuf:2008bb}). 
Then for all points  $Q^2>Q_{\nu}^2$ and $\tau_{\text{rc2}}$ decreases with rising $\nu$.

\begin{figure}
\vspace{0.5cm}
\centering
\includegraphics[width=7cm,angle=0]{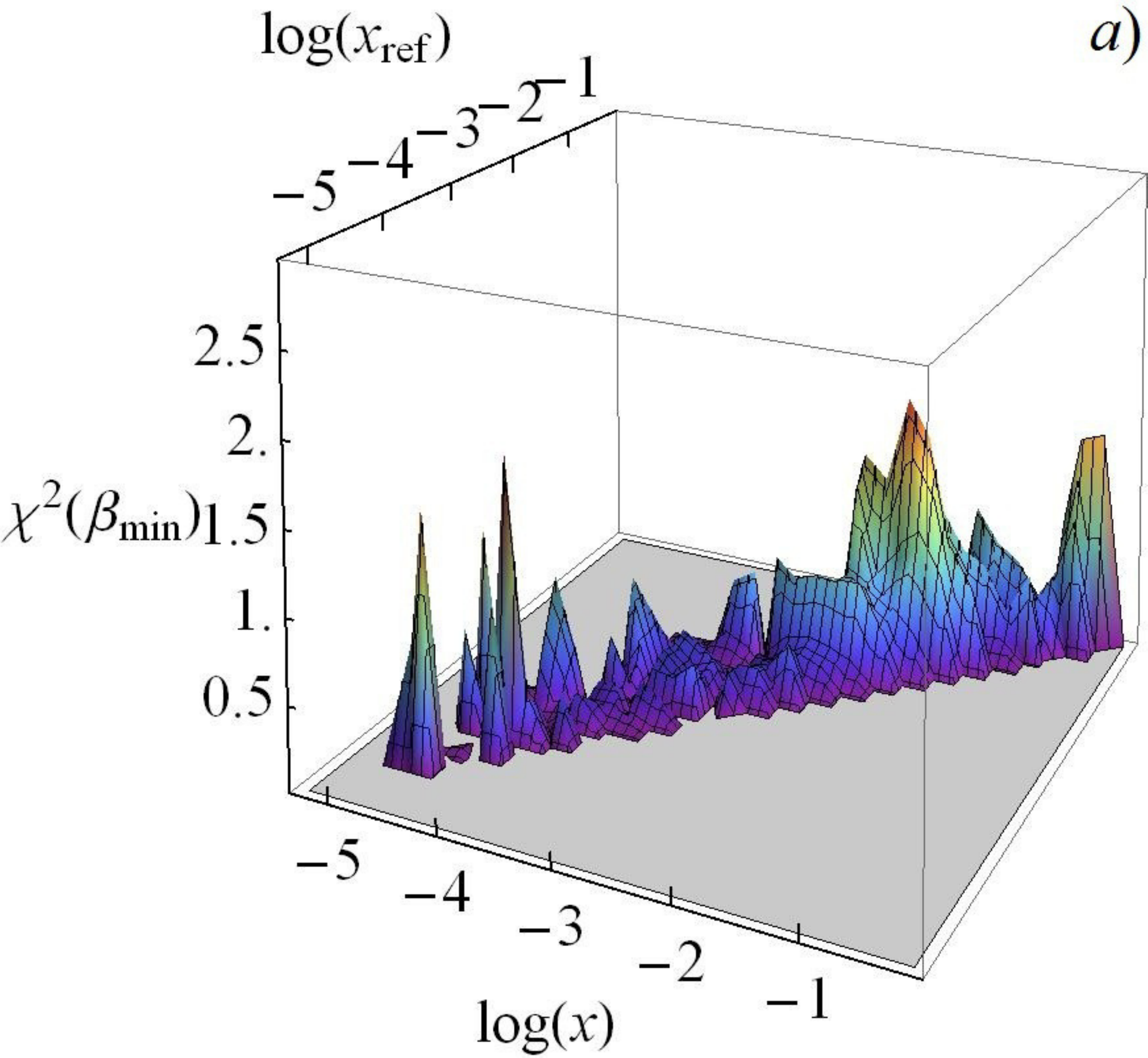}~~~
\includegraphics[width=7cm,angle=0]{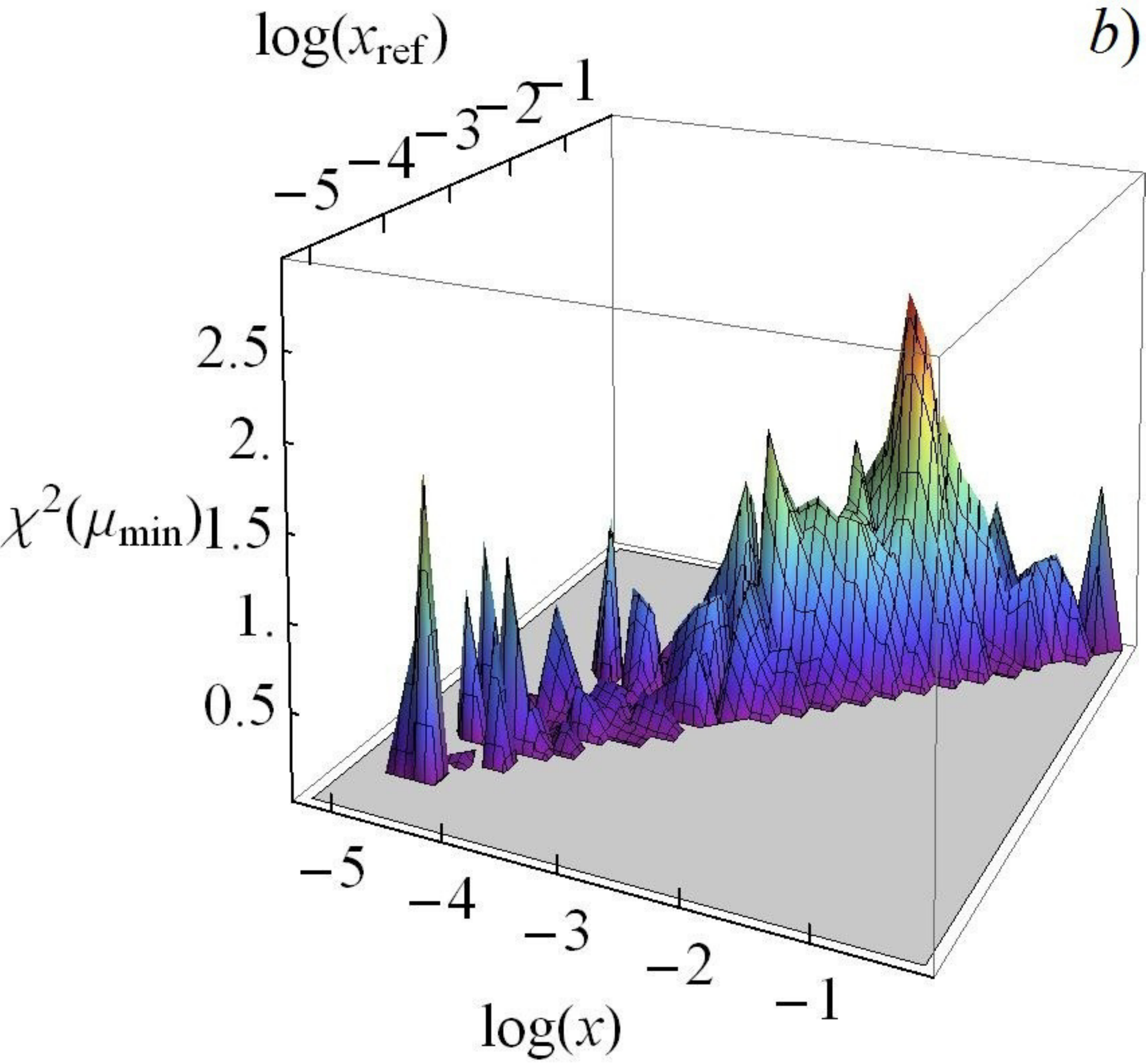} \\%
\includegraphics[width=7cm,angle=0]{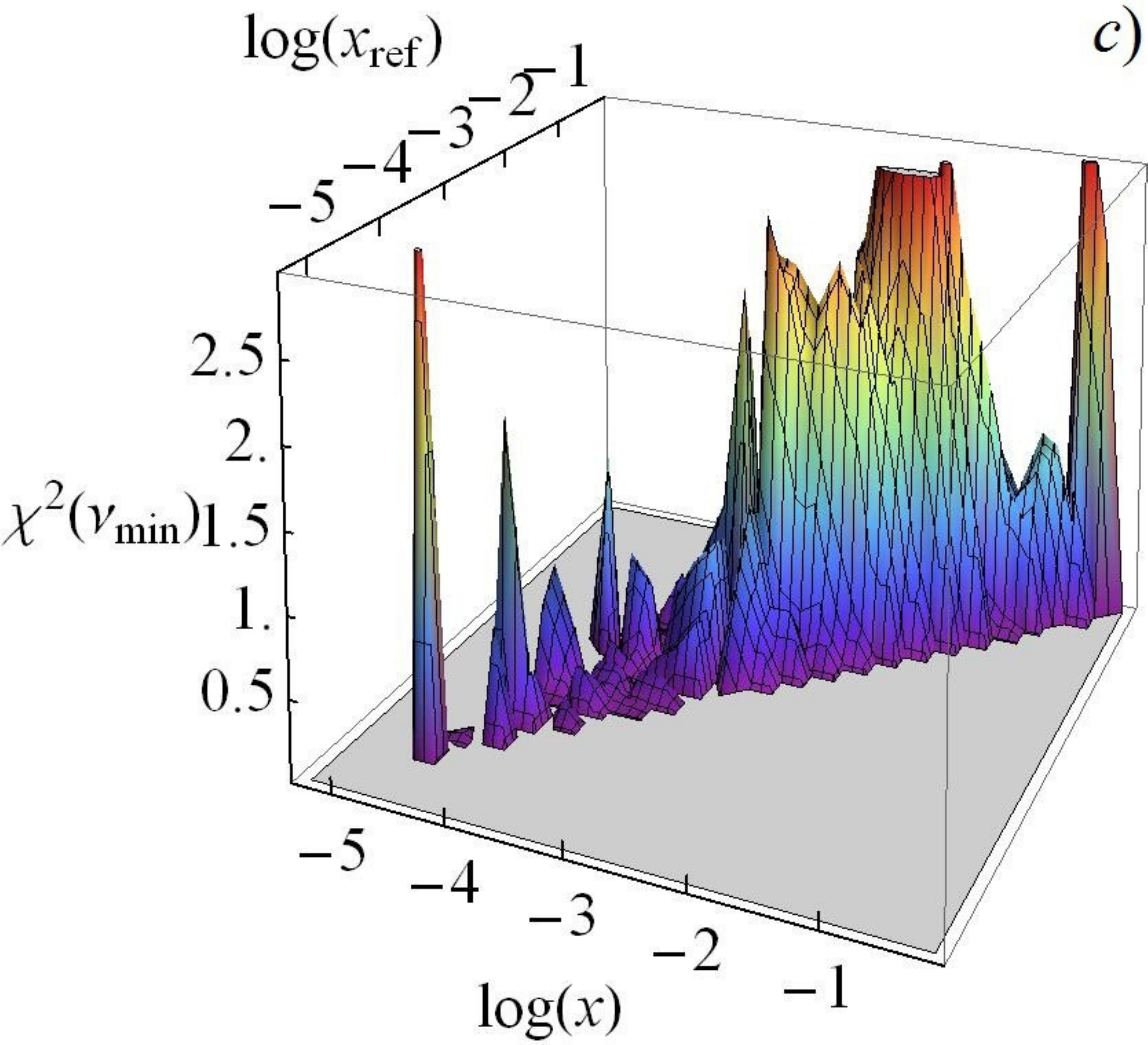}~~~
\includegraphics[width=7cm,angle=0]{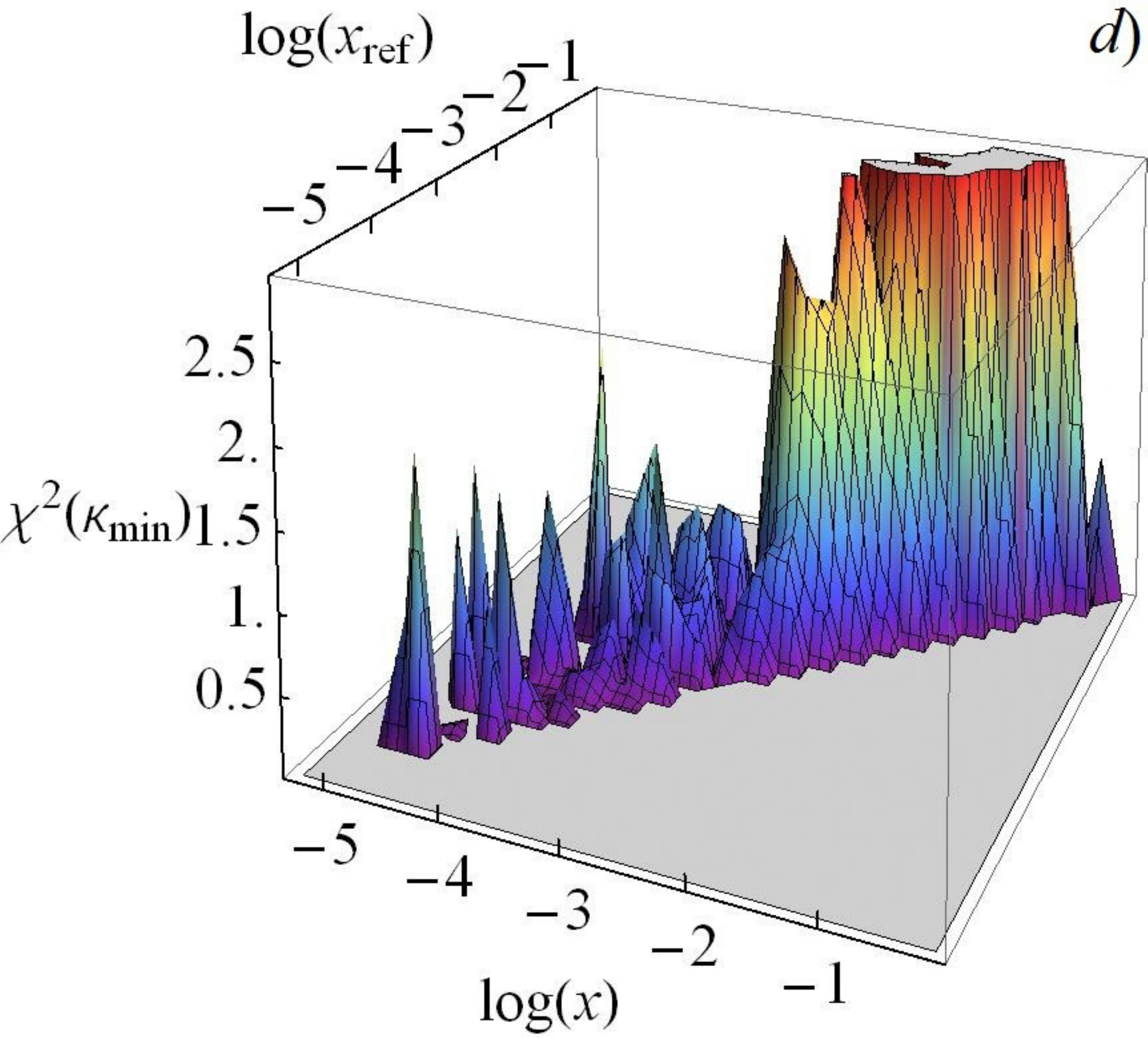}\\
\caption{Three dimensional plots of minimal values of $\chi^2$
functions~(\ref{chix1}) for different scaling variables:
a) logarithmic scaling variable $\tau_{\text{phn}}$ (\ref{eff}),
running coupling b) rc1 (\ref{rc1}), c) rc2 (\ref{rc2}) and
d) diffusive scaling (\ref{ds}).}%
\label{contplots}%
\end{figure}

Let us first examine 3 dimensional plots of $\alpha_{\text{min}%
}(x,x_{\text{ref}})$ (note again that $\alpha=\beta,\,\mu,\,\nu$ or $\kappa,$
depending on the scaling variable). For GS to hold there should be a visible
plateau of $\alpha_{\text{min}}$ over some relatively large part of
$(x,x_{\text{ref}})$ space (recall that by construction $x<x_{\text{ref}}$).
Looking at Fig.~\ref{3dplots} one has to remember that the values of
$\alpha_{\text{min}}(x,x_{\text{ref}})$ are subject to fluctuations that will
be "averaged over" when we discuss more "integrated" quantities $\left\langle
\alpha_{\text{min}}\right\rangle $ and $\left\langle \left\langle
\alpha_{\text{min}}\right\rangle \right\rangle $. Note that statistical
errors of  $\alpha_{\text{min}}(x,x_{\text{ref}})$ which are quite large for small $x$
are not displayed in Fig.~\ref{3dplots}. One can can conclude from
Fig.~\ref{3dplots} that for all 4 cases (\ref{eff}) -- (\ref{ds}) there is
rather strong dependence of $\alpha_{\text{min}}(x,x_{\text{ref}})$ for large
values of $x$ and $x_{\text{ref}}$. In the case of $Q^{2}$-dependent scaling
variable (\ref{eff}) (Fig.~\ref{3dplots}.a) and for the running coupling case (\ref{rc1},\ref{rc2}) 
(Figs.~\ref{3dplots}.b, c),
the values of parameters $\beta$, $\mu$ and $\nu$ rise steeply for large
$x$'s, whereas for diffusive scaling parameter $\kappa$ is falling down
rapidly. More closer look reveals that for running coupling rc1 case 
(Fig.~\ref{3dplots}.b) there is in fact no distinct plateau, one can also see a
systematic rise of $\mu_{\text{min}}$ in a region of very small $x$'s.
Similarly for the diffusive scaling (Fig.~\ref{3dplots}.d) we see rather
systematic growth of $\kappa_{\text{min}}$ for small $x$'s with possible
plateau in a small corner of very low $x$'s. At first glance no plateau is neither 
seen for $\beta_{\text{min}}(x,x_{\text{ref}})$ (Fig.~\ref{3dplots}.a). 
However -- as will be shown in the following  -- because of considerable 
statistical uncertainties within the scale used in Fig.~\ref{3dplots}.a, 
very good  description of GS with constant $\beta$ is still possible.

It is interesting to look at 3 dimensional  plots of the corresponding $\chi^{2}$
values (\ref{chix1}) shown in Fig.~\ref{contplots}. Recal that for GS to hold one should
observe small values of $\chi^{2}(\alpha_{\text{min}})$ in the same region where
$\alpha_{\text{min}}$ is constant.  This happens for $\tau_{\rm phn}$ (Fig.~\ref{contplots}.a)
where  $\chi^{2}$ oscillates around 1 not exceeding 2 even for large values of $x$.
Similarly $\tau_{\text{rc1}}$ (Fig.~\ref{contplots}.b)  stays smaller than
 2 up to $x \sim 10^{-2}$ where $\chi^{2}$ jumps above 2. In this region, however,
parameter $\mu$ is steadily decreasing with $x$.
In contrast, in the case of $\tau_{\text{rc2}}$ (Fig.~\ref{contplots}.c) and
$\tau_{\text{ds}}$ $\chi^{2}$ (Fig.~\ref{contplots}.d) $\chi^2$
have pronounced fluctuations and a plateau (if at all) is visible only below
$x \sim 10^{-3}$.  However, in this region parameter $\nu$ (corresponding to Fig.~\ref{contplots}.c)
rises with $x$, whereas $\kappa$ (corresponding to Fig.~\ref{contplots}.d)
exhibits rather strong fluctuations.

\begin{figure}[ptb]
\centering
\includegraphics[width=7cm,angle=0]{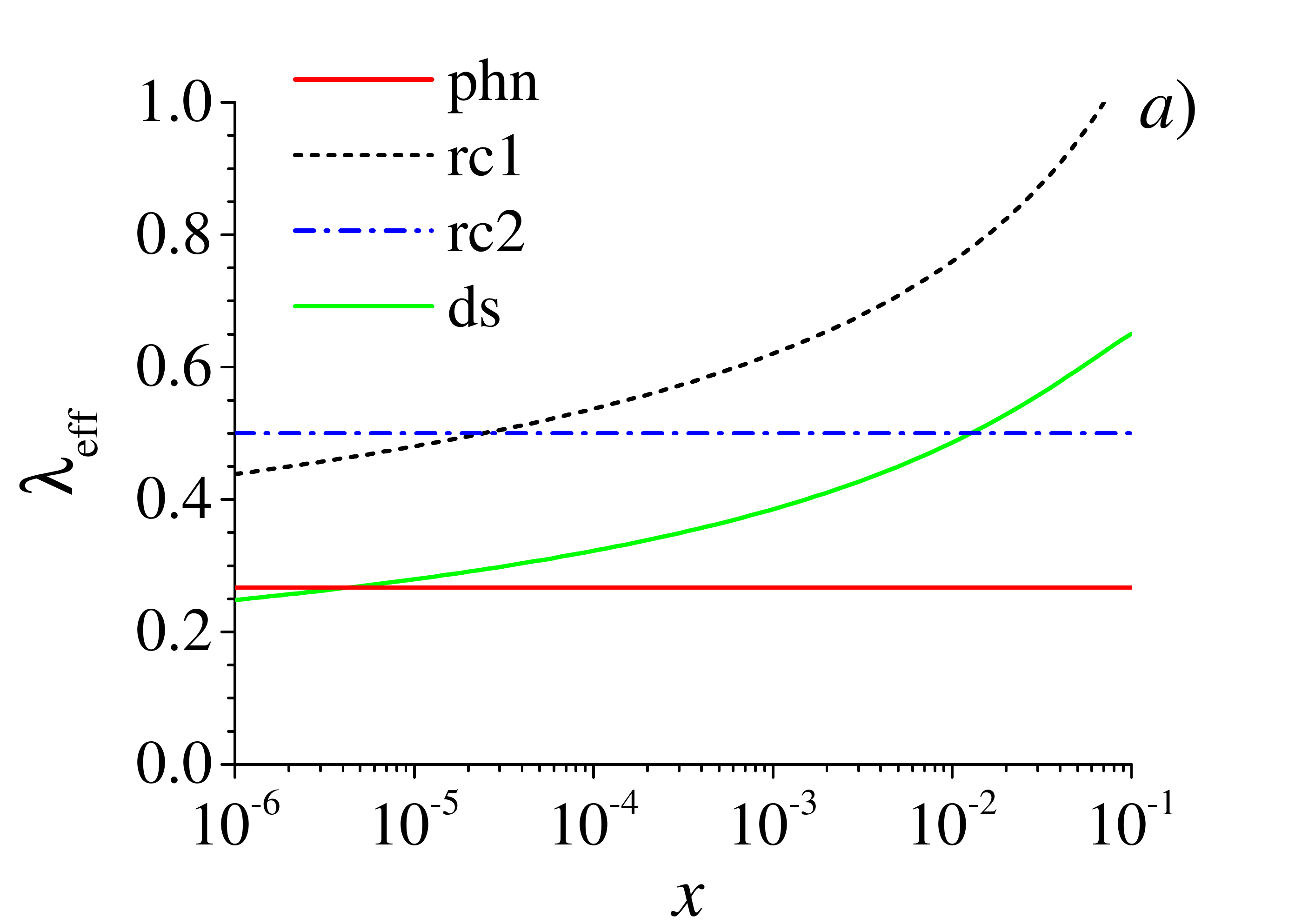}~~~
\includegraphics[width=7cm,angle=0]{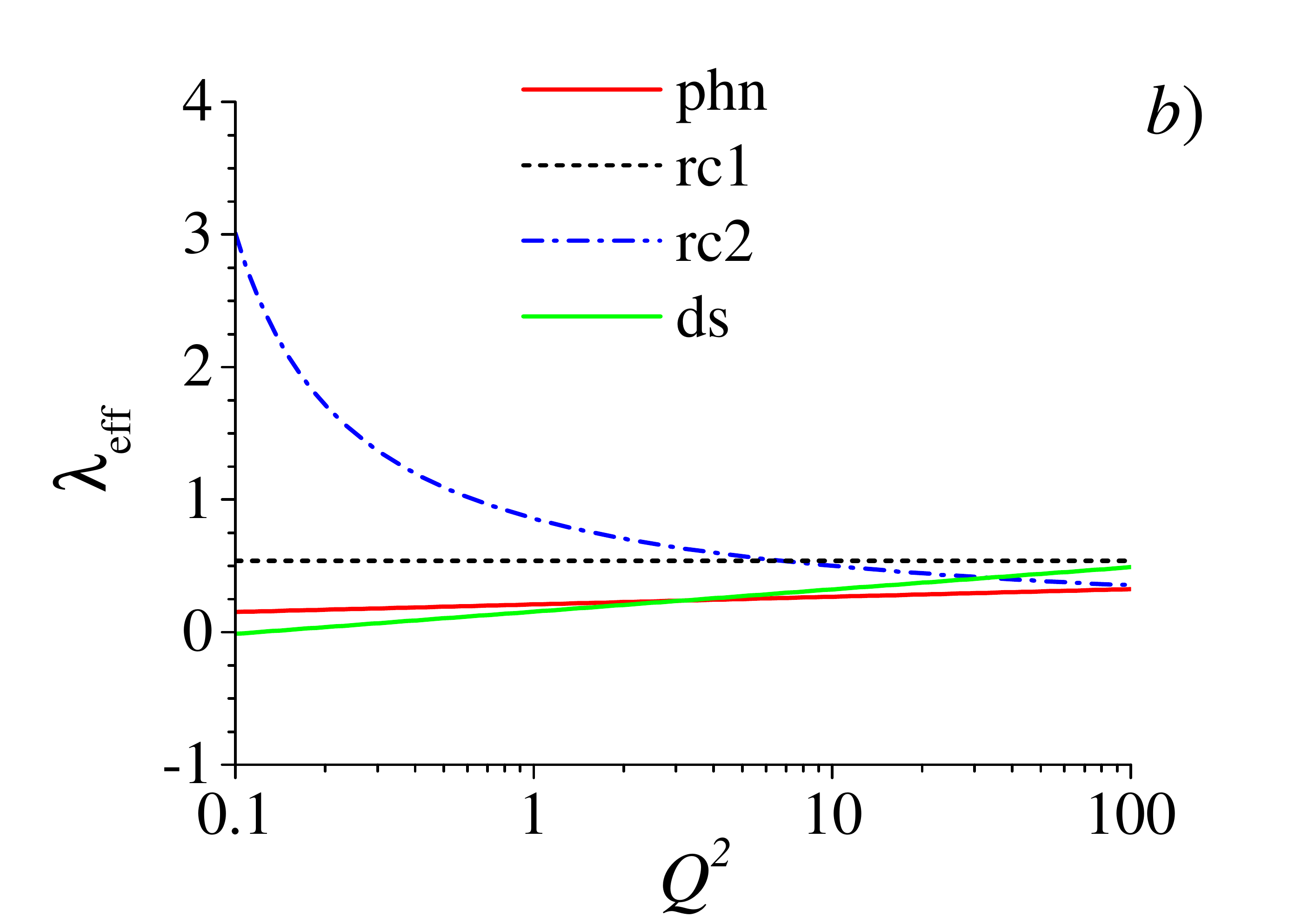}\caption{Effective exponents
(\ref{leffdef}) as functions of $x$ for fixed $Q^{2}=10$~GeV$^{2}$/$c^{2}$
(left) and as functions of $Q^{2}$ for fixed $x=0.0001$ (right).}%
\label{leffxQ}%
\end{figure}

\begin{figure}
\centering
\includegraphics[width=7cm,angle=0]{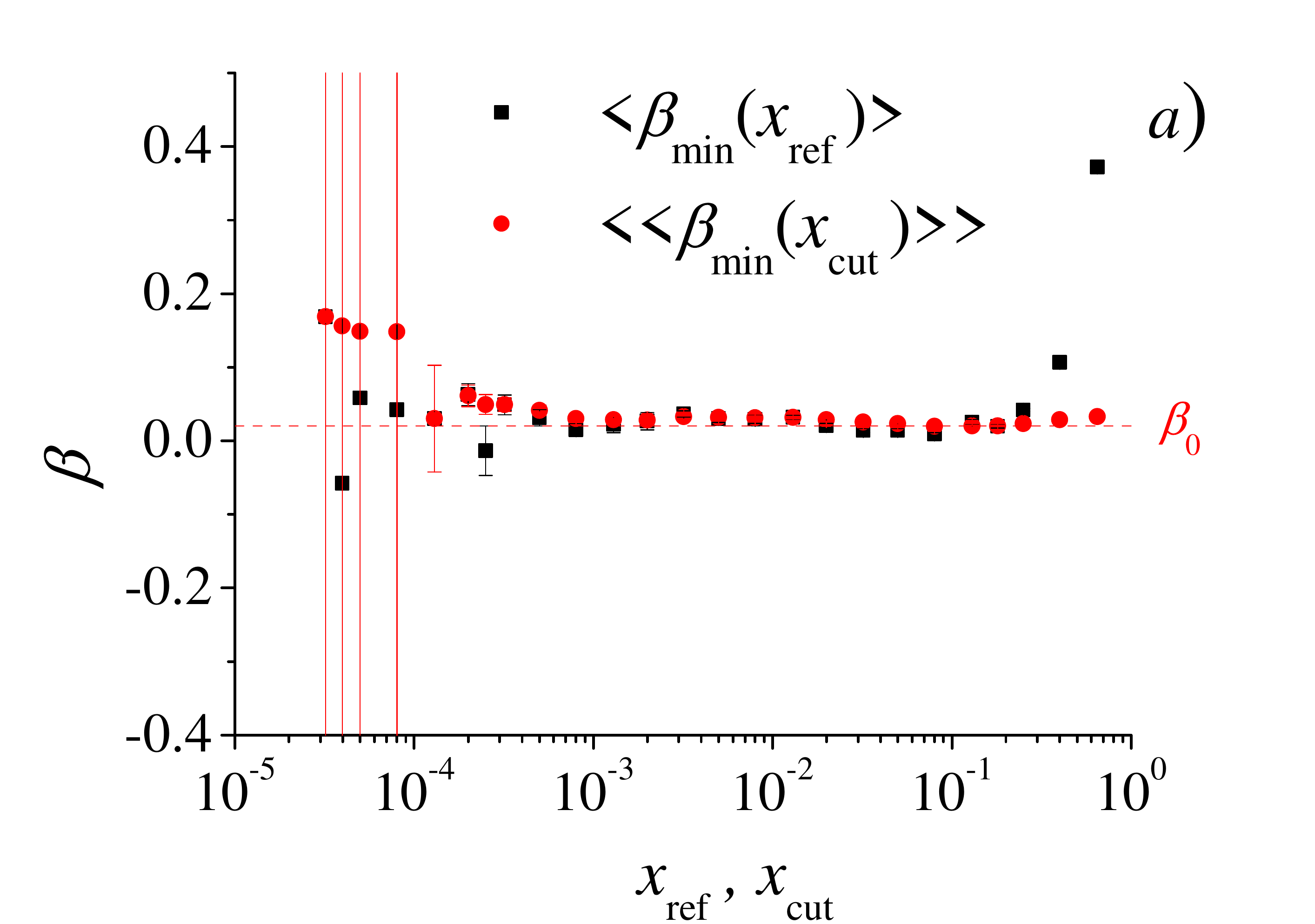}
\includegraphics[width=7cm,angle=0]{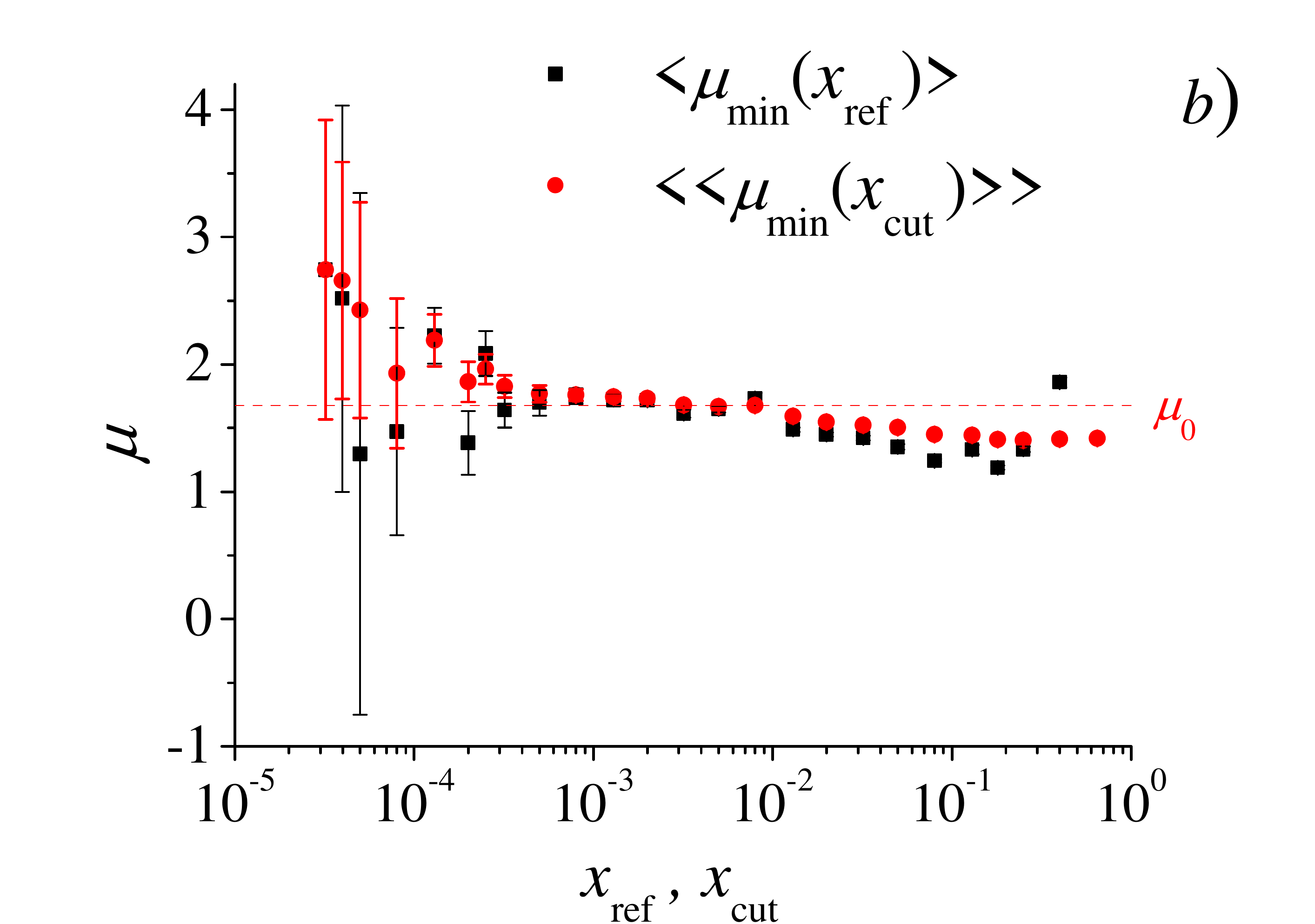} \\%
\vspace{0.5cm}
\includegraphics[width=7cm,angle=0]{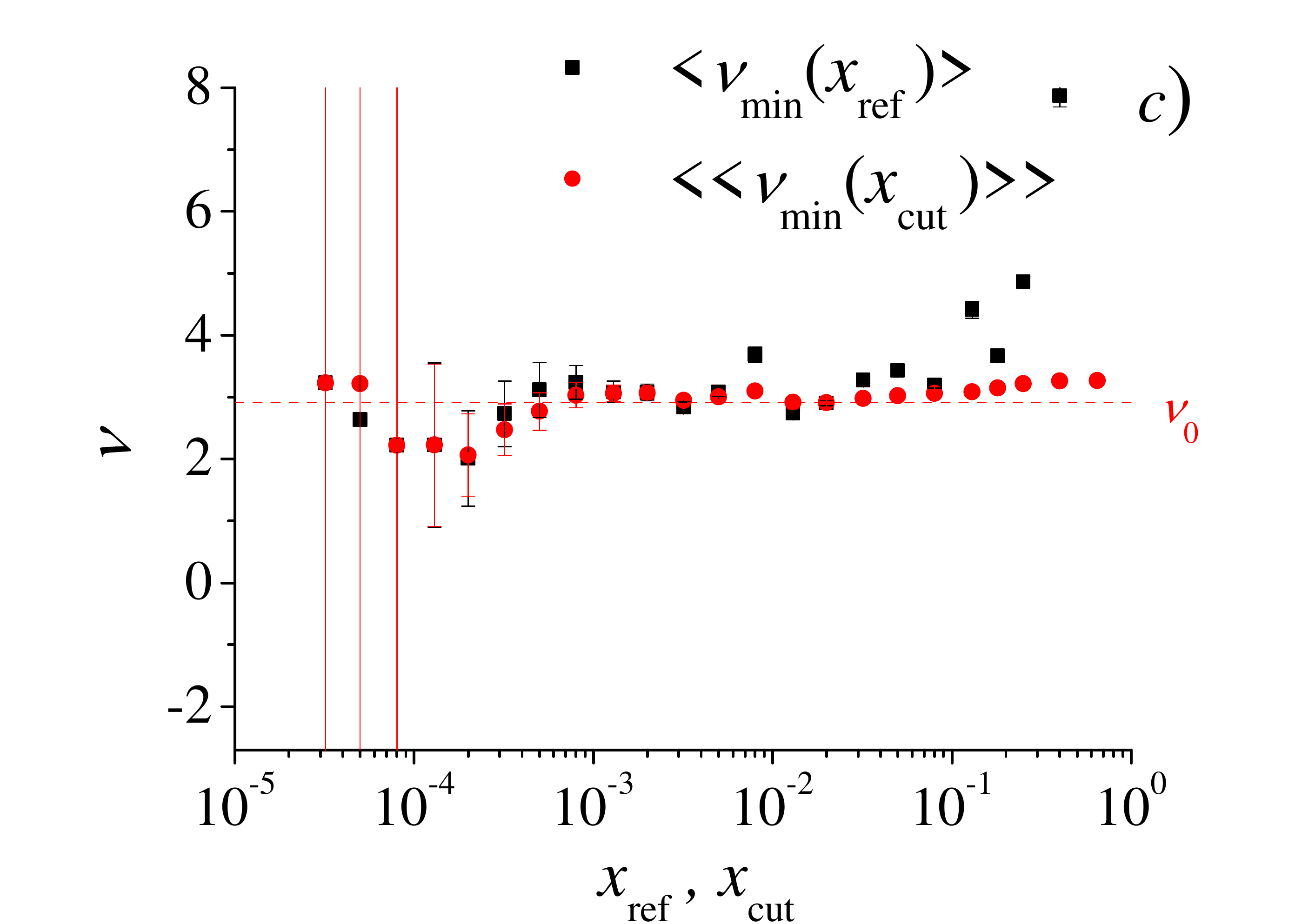}
\includegraphics[width=7cm,angle=0]{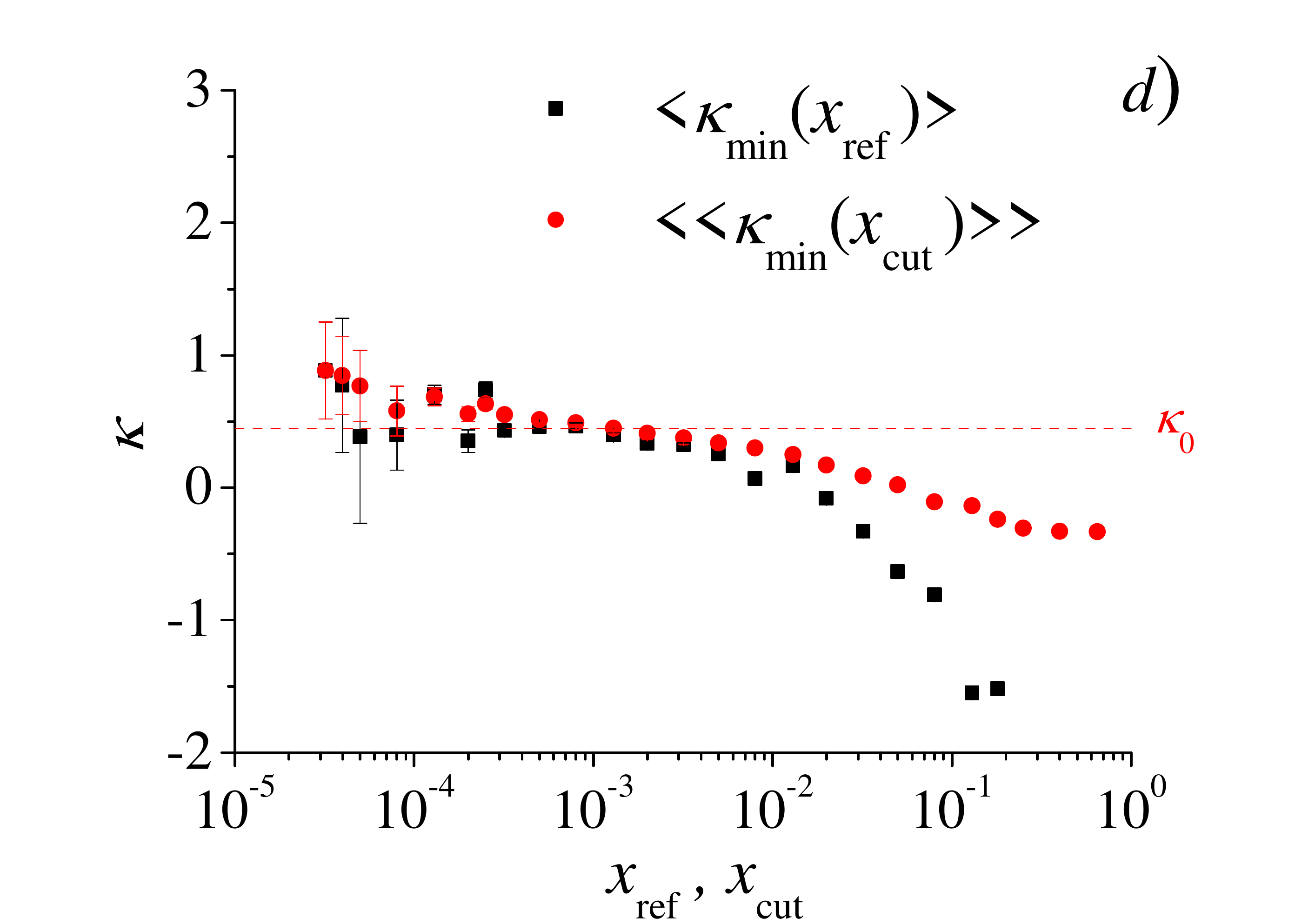}
\caption{Averaged
values $\left\langle \alpha_{\mathrm{min}}(x_{\mathrm{ref}})\right\rangle $
(black squares) and $\left\langle \left\langle \alpha_{\mathrm{min}%
}(x_{\mathrm{cut}})\right\rangle \right\rangle $ (red circles) for different
scaling hypotheses: a)  logarithmic $Q^2$ effective exponent (\ref{eff}) with
$\alpha=\beta$, running coupling
scaling variables b) rc1 (\ref{rc1}) with $\alpha=\mu$ and d) rc2 (\ref{rc2})
with $\alpha=\nu$, and d) diffusive scaling (\ref{ds}) with $\alpha=\kappa$, 
respectively.}
\label{lamplots}%
\end{figure}

Due to different functional dependence of the saturation scales entering
Eqs.~(\ref{eff}) -- (\ref{ds}) variations of
parameters $\beta$, $\mu$, $\nu$ and $\kappa$ differently influence
pertinent scaling variable $\tau$. Therefore --
before we turn to average quantities $\left\langle \ldots\right\rangle $ and
$\left\langle \left\langle \ldots\right\rangle \right\rangle $ displayed in
Fig.~\ref{lamplots} -- let us  define 
{\em effective} exponents
$\lambda_{\mathrm{eff}}$:
\begin{equation}
\lambda_{\mathrm{eff}}(x,Q^{2})=\log\big(\frac{\tau}{Q^{2}}\big)/\log(x)
\label{leffdef}%
\end{equation}
which depend on fitting parameters $\beta,\,\mu,\,\nu$ and $\kappa$. In
Fig.~\ref{leffxQ} we plot these effective powers as functions of $x$ and
$Q^{2}$ for the values of the parameters 
$\beta_{\rm min},\,\mu_{\rm min},\,\nu_{\rm min}$ 
and $\kappa_{\rm min}$
fixed at the end of this Section.

In order to find the scale relevant for a parameter entering definition of a
given scaling variable $\tau$ (\ref{eff}) -- (\ref{ds}), for each scaling
hypothesis separately we have varied this parameter around the {\em best} 
value by $\pm\epsilon$ and required that
\begin{equation}
|\lambda_{\mathrm{eff}}(\alpha_{\rm min}\pm\epsilon;x,Q^{2})-\lambda_{\mathrm{eff}%
}(\alpha_{\rm min};x,Q^{2})|=1
\end{equation}
for some typical values of $x=0.0001$ and $Q^{2}=10$~GeV$^{2}/c^{2}$. In this
way in each case the value of $\epsilon$ provides the reference scale for each 
variational parameter $\alpha=\beta,\,\mu,\,\nu$ or $\kappa$. 
Therefore looking at Fig.~\ref{lamplots} one
should bear in mind that the span of the vertical axis corresponds to the
variation of the effective exponent $\Delta\lambda_{\mathrm{eff}}\sim\pm1$
around its {\em best} value.

Looking at Figs.~\ref{lamplots} we see immediately that the best scaling
properties are exhibited by parameter $\beta$ of $Q^{2}$-dependent scaling
variable $\tau_{\rm phn}$ (\ref{eff}). Parameter $\beta$ is well described by a
constant
\begin{equation}
\beta_{0}=\langle\langle\beta_{\mathrm{min}}(0.08)\rangle\rangle=0.02\pm0.001
\label{beta0}%
\end{equation}
over 3 orders of magnitude in $x$. We have used the value of maximal
$x_{\text{cut}}=0.08$, since it was the value of $x_{\text{cut}}$ for which
$\lambda_{0}=0.329$ has been extracted in Ref.~\cite{Praszalowicz:2012zh}, 
although -- as clearly seen from
Fig.~\ref{lamplots}.a -- GS in variable $\tau_{\rm phn}$ works well up to
$x\simeq0.2$. There is an impressive agreement between both averages
$\langle\beta_{\mathrm{min}}\rangle$ and $\langle\langle\beta_{\mathrm{min}%
}\rangle\rangle,$ however the value (\ref{beta0}) is five times smaller than
expected from the fit to low $x$ behavior of $F_{2}$ structure function
(\ref{logfit}).

\begin{figure}
\vspace{0.5cm}
\centering
\includegraphics[width=7cm,angle=0]{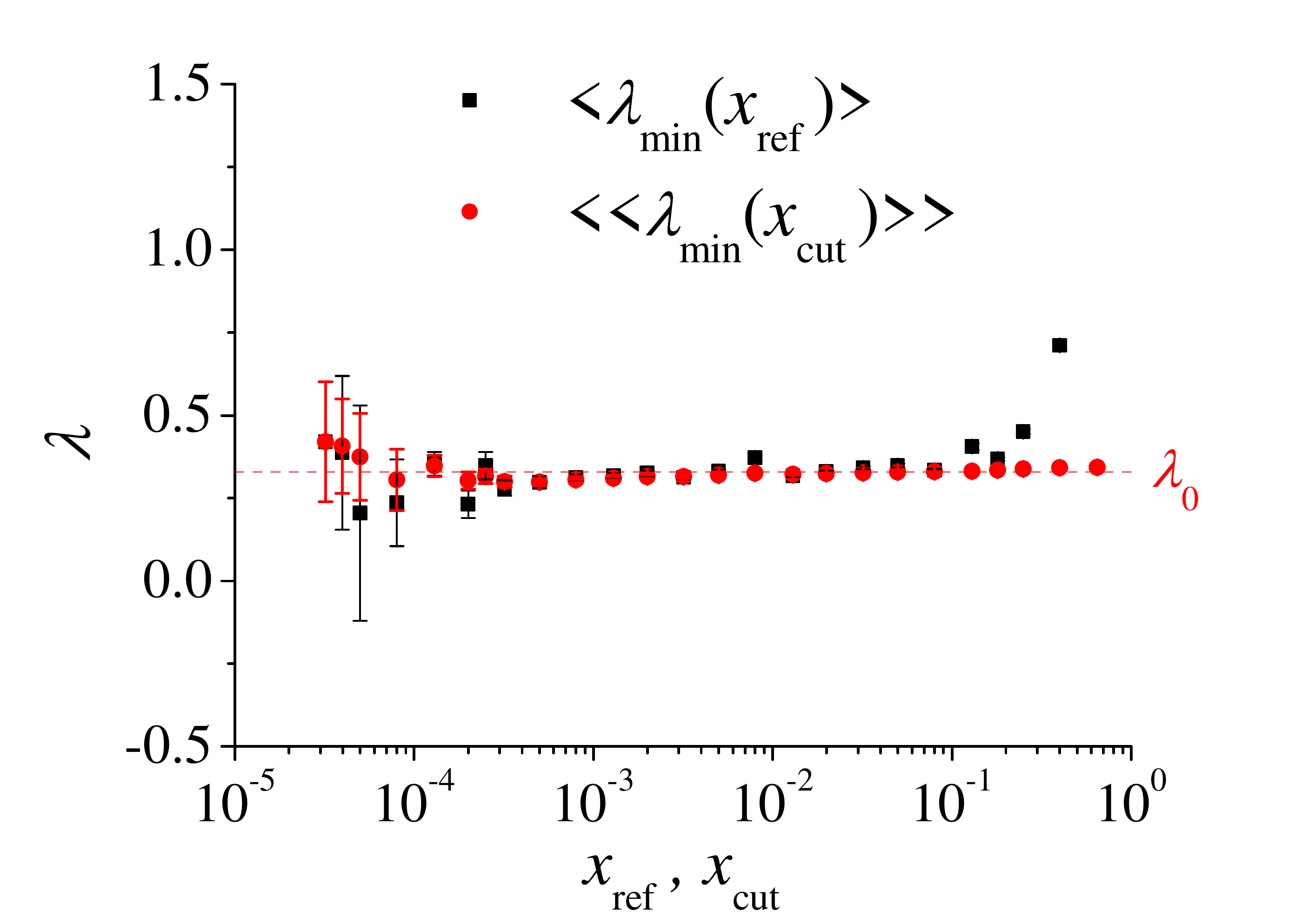}
\caption{Averaged
values $\left\langle \lambda_{\mathrm{min}}(x_{\mathrm{ref}})\right\rangle $
(black squares) and $\left\langle \left\langle \lambda_{\mathrm{min}%
}(x_{\mathrm{cut}})\right\rangle \right\rangle $ (red circles) for scaling
variable with constant exponent $\lambda$ (Eqs.~(\ref{tau}) and (\ref{Qsat})).
Figure from Ref.~\cite{Praszalowicz:2012zh}.}%
\label{lamconst}%
\end{figure}

For comparison in Fig.~\ref{lamconst} we present the plot from Ref.~\cite{Praszalowicz:2012zh}
where $\langle\lambda_{\mathrm{min}}\rangle$ and $\langle\langle
\lambda_{\mathrm{min}}\rangle\rangle$ for scaling hypothesis with constant
$\lambda$ (\emph{i.e. }for\emph{ }$\beta=0$) are shown. We see that the
quality of a fit with a constant $\lambda$ is only a little worse than GS in
$\tau_{\rm phn}$ but in general much better than in the case of the remaining
scaling variables (\ref{rc1}) -- (\ref{ds}).

Indeed, for the running coupling constant rc1 case (\ref{rc1}) we see in
Fig.~\ref{lamplots}.b monotonous fall of $\langle\mu_{\mathrm{min}}\rangle$
and $\langle\langle\mu_{\mathrm{min}}\rangle\rangle$ with $x_{\text{ref}}$ and
$x_{\text{cut}}$ respectively, although the large errors at small $x$'s allow
for a constant fit up to $x_{\text{ref}}$, $x_{\text{cut}}\simeq0.008$
yielding%
\begin{equation}
\mu_{0}=\langle\langle\mu_{\mathrm{min}}(0.008)\rangle\rangle=1.677\pm0.014.
\label{mu0}%
\end{equation}
The situation is similar for running coupling rc2 case (\ref{rc2}) where the
constant fit is possible up to $x_{\text{ref}}$, $x_{\text{cut}}\simeq0.02$
(see Fig.~\ref{lamplots}.c) giving%
\begin{equation}
\nu_{0}=\langle\langle\nu_{\mathrm{min}}(0.02)\rangle\rangle=2.909\pm0.025.
\label{nu0}%
\end{equation}
In this case, however, one should bear in mind that more "differential" measure
of GS - $\chi^{2}(\nu_{\mathrm{min}})$ - shown in Fig.~\ref{contplots}.c does not support hypothesis of GS above
$x\sim 10^{-3}$.

Finally, in the case of diffusive scaling (\ref{ds}) we can hardly conclude
that GS is really seen; although it is possible to find constant behavior of
$\langle\kappa_{\mathrm{min}}\rangle$ and $\langle\langle\kappa_{\mathrm{min}%
}\rangle\rangle$ below $x\sim10^{-3}$ with%
\begin{equation}
\kappa_{0}=\langle\langle\kappa_{\mathrm{min}}(0.0013)\rangle\rangle
=0.449\pm0.012. \label{kappa0}%
\end{equation}

Note, that the errors in Eqs.~(\ref{beta0}) -- (\ref{kappa0})
are purely statistical (for discussion of systematic uncertainties see \cite{Praszalowicz:2012zh}). 

\section{Summary and Conclusions}
\label{sumcon}

In this paper we have applied the method developed in Refs.\cite{Stebel:2012ky,Praszalowicz:2012zh} 
to assess the quality
of geometrical scaling of $e^+p$ DIS data on $F_2$ as provided by the combined
H1 and ZEUS analysis of Ref.~\cite{HERAcombined}. In a sense our analysis is in a spirit of previous works
\cite{Gelis:2006bs,Beuf:2008bb} and especially Ref.~\cite{Royon:2010tz} where the same set
of data has been analyzed by means of so called quality factor. 
Although the
authors of Ref.~\cite{Royon:2010tz} applied kinematical cuts $4\leq Q^2 \leq 150 \text{GeV}^2$, $x \leq 0.01$
our results for scaling parameters given in Eqs.~(\ref{lambda0}) 
and (\ref{mu0})~--~(\ref{kappa0}) are in good agreement
with their findings.
 For completeness let us quote their results (note
that they did not consider logarithmic $Q^2$ dependence of $\tau_{\rm phn}$):  $\mu_0=1.61$ (rc1), $\nu_0=2.76$ (rc2) and $\kappa_0=0.31$ (ds). 
Difference in $\kappa_0$ can be explained by applied kinematical cuts, indeed, if we take  maximal $x_{\rm cut}=0.01$ we obtain 
$\langle\langle\kappa_{\mathrm{min}}(0.01)\rangle\rangle=0.301\pm0.006$  in agreement with \cite{Royon:2010tz}.

Despite the fact that we have been able to find some corners of phase space where
geometrical scaling in variables (\ref{rc1})~--~(\ref{ds}) could be seen, it is
absolutely clear that the best scaling variable is given by (\ref{eff}) (or even by a constant
$\lambda$ of Eq.~(\ref{lambda0})), whereas {\em diffusive scaling} hypothesis is
certainly ruled out. 
This is quite well illustrated  in Fig.~\ref{leffxQ} where effective
exponent $\lambda_{\rm eff}$ for scaling variable (\ref{ds}) changes sign for small
$Q^2$.  This is the reason why in Ref.~\cite{Royon:2010tz} a cut on low $Q^2$ has been
applied. Similar argument applies for the running coupling rc2 case (\ref{rc2}) which blows 
 up for small $Q^2$. Because of that $\chi_{x_{i},x_{\text{ref}}}^{2}$ functions have no minima 
 for very low $x_{i}$ and $x_{\text{ref}}$ (points with small $x$ have also small $Q^2$).  
 Therefore the only candidate for scaling variable is running coupling rc1 case (\ref{rc1}). Nevertheless, 
 comparing Fig.~\ref{lamplots}.b with Fig.~\ref{lamconst} where we plot results for
 GS scaling with constant exponent $\lambda$, we see that both by quality and applicability
 range, the original  form of scaling variable does much better job than (\ref{rc1}). Although
 our results for best values of parameters entering definitions of scaling variables  (\ref{rc1})~--~(\ref{ds})
are in agreement with Refs.~\cite{Gelis:2006bs,Beuf:2008bb,Royon:2010tz} we do not
confirm their conclusion that only diffusive scaling is ruled out while for other forms of scaling variable
geometrical scaling is of similar quality. 
It is of course perfectly possible that the HERA data are not "enough asymptotic" and 
geometrical scaling in one of the variables defined in Eqs.~(\ref{rc1})~--~(\ref{ds}) will 
show up at higher energies and lower Bjorken $x$'s.

\section*{Acknowledgements}

MP would like to thank Robi Peschanski for discussion and for drawing his
attention to the quality factor studies of geometrical scaling . 
This work was supported by the Polish NCN 
grant 2011/01/B/ST2/00492.


\begin{thebibliography}{99} 


\bibitem {Stasto:2000er}A.M.~Stasto, K.J.~Golec-Biernat, J.~Kwiecinski,
\emph{Geometric scaling for the total $\gamma^{*}p$ cross-section in the low x
region}, \emph{Phys.\ Rev.\ Lett.}\ \textbf{86} (2001) 596 [arXiv:hep-ph/0007192].



\bibitem {GolecBiernat:1998js}K.J.~Golec-Biernat, M.~Wusthoff,
\emph{Saturation effects in deep inelastic scattering at low $Q^{2}$ and its
implications on diffraction}, \emph{Phys.\ Rev.\ D} \textbf{59} (1998) 014017
[arXiv:hep-ph/9807513];\newline
\emph{Saturation in diffractive deep inelastic
scattering}, \emph{Phys.\ Rev.\ D }\textbf{60} (1999) 114023
[arXiv:hep-ph/9903358].


\bibitem{Praszalowicz:2012zh}
  M.~Praszalowicz and T.~Stebel,
\emph{Quantitative Study of Geometrical Scaling in Deep Inelastic Scattering at HERA},
 [arXiv:1211.5305 [hep-ph]].
  
  \bibitem{Stebel:2012ky}T.~Stebel, Master Thesis, \emph{Quantitative analysis
of Geometrical Scaling in Deep Inelastic Scattering}, [arXiv:1210.1567
[hep-ph]].

\bibitem {Mueller:2001fv}A.~H.~Mueller, \emph{Parton Saturation: An Overview},
[arXiv:hep-ph/0111244].


\bibitem {McLerran:2010ub}L.~McLerran, \emph{Strongly Interacting Matter
Matter at Very High Energy Density: 3 Lectures in Zakopane}, \emph{Acta
Phys.\ Pol.\ B} \textbf{41} (2010) 2799 [arXiv:1011.3203 [hep-ph]].

\bibitem{jimwlk}
J. Jalilian-Marian, A. Kovner, A. Leonidov and H. Weigert, 
\emph{The BFKL equation from the Wilson renormalization group, Nucl. Phys. B} {\bf 504} (1997) 415 [arXiv:hep-ph/9701284];\\ 
\emph{The Wilson renormalization group for low x physics: Towards the high density regime, Phys. Rev. D} {\bf 59} (1999) 014014 [arXiv:hep-ph/9706377];\\
E. Iancu, A. Leonidov and L. D. McLerran, 
\emph{Nonlinear gluon evolution in the color glass condensate. 1. Nucl. Phys. A} {\bf 692} (2001) 583 [hep-ph/0011241];\\ 
E. Ferreiro, E. Iancu, A. Leonidov and L. D. McLerran, 
\emph{Nonlinear gluon evolution in the color glass condensate. 2. Nucl. Phys. A} {\bf 703} (2002) 489
[hep-ph/0109115]. 

\bibitem{BK}
  I.~Balitsky,
\emph{Operator expansion for high-energy scattering},
\emph{ Nucl.\ Phys.\ B} {\bf 463} (1996) 99
  [hep-ph/9509348];\newline
  Y.~V.~Kovchegov,
\emph{Small x} $F_2$ \emph{ structure function of a nucleus including multiple pomeron exchanges},
\emph{Phys.\ Rev.\ D} {\bf 60} (1999) 034008
  [hep-ph/9901281];\newline
    Y.~V.~Kovchegov,
\emph{Unitarization of the BFKL pomeron on a nucleus},
 \emph{Phys.\ Rev.\ D} {\bf 61} (2000) 074018
  [hep-ph/9905214].

\bibitem{Munier:2003vc}
  S.~Munier and R.~B.~Peschanski,
 \emph{Geometric scaling as traveling waves},
 \emph{Phys.\ Rev.\ Lett.}  {\bf 91} (2003) 232001
  [hep-ph/0309177];\newline
 S.~Munier and R.~B.~Peschanski,
\emph{Traveling wave fronts and the transition to saturation},
  \emph{Phys.\ Rev.\ D} {\bf 69} (2004) 034008
  [hep-ph/0310357].
 

\bibitem{sat1}
L. V. Gribov, E. M. Levin and M. G. Ryskin,
\emph{Semihard Processes In QCD},
\emph{Phys.\ Rept.}\ \textbf{100} (1983) 1;\newline
A. H. Mueller and  J-W. Qiu, 
\emph{Gluon recombination and shadowing at small values of x},
Nucl. Phys. {\bf 268}, 427(1986);\newline
 A. H. Mueller,
\emph{Parton Saturation at Small x and in Large Nuclei},
 Nucl. Phys. {\bf B558}, 285 (1999) [arXiv:hep-ph/9904404]. 

\bibitem{MLV} 
  L.~D.~McLerran and R.~Venugopalan,
\emph{Computing quark and gluon distribution functions for very large nuclei},
\emph{Phys.\ Rev.\ D} \textbf{49} (1994) 2233 [arXiv:hep-ph/9309289]; \\
\emph{Gluon distribution functions for very large nuclei at small transverse
momentum},
\emph{Phys.\ Rev.\ D} \textbf{49} (1994) 3352 [arXiv:hep-ph/9311205] ;\\
\emph{Green's functions in the color field of a large nucleus},
\emph{Phys.\ Rev.\ D} \textbf{50} (1994) 2225 [arXiv:hep-ph/9402335].
  
  \bibitem{DGLAP}
  V.~N.~Gribov and L.~N.~Lipatov,
 \emph{Deep inelastic e p scattering in perturbation theory},
 \emph{Sov.\ J.\ Nucl.\ Phys.}  {\bf 15} (1972) 438
  (\emph{Yad.\ Fiz.}  {\bf 15} (1972) 781);\newline
    G.~Altarelli and G.~Parisi,
\emph{Asymptotic Freedom in Parton Language},
\emph{ Nucl.\ Phys.\ B} {\bf 126} (1977) 298;\newline
Y.~L.~Dokshitzer,
\emph{Calculation of the Structure Functions for Deep Inelastic Scattering and} $e^+ e^-$ 
\emph{Annihilation by Perturbation Theory in Quantum Chromodynamics},
\emph{Sov.\ Phys.\ JETP} {\bf 46} (1977) 641
 (\emph{Zh.\ Eksp.\ Teor.\ Fiz.}  {\bf 73} (1977) 1216).
 
\bibitem{Kwiecinski:2002ep} 
  J.~Kwiecinski and A.~M.~Stasto,
\emph{Geometric scaling and QCD evolution},
   \emph{Phys.\ Rev.\ D} {\bf 66}, 014013 (2002) and
\emph{Large geometric scaling and QCD evolution},
   \emph{Acta Phys.\ Polon.\ B} {\bf 33}, 3439 (2002).
 
 
  
\bibitem{BFKL}
E.A. Kuraev, L.N. Lipatov, V.S. Fadin,
\emph{The Pomeranchuk Singularity in Nonabelian Gauge Theories},
\emph{Sov. Phys. JETP} 45 (1977) 199
(\emph{Zh. Eksp. Teor. Fiz.} 72 (1977) 377);\newline
I.I. Balitsky, L.N. Lipatov,
\emph{The Pomeranchuk Singularity in Quantum Chromodynamics},
\emph{Sov. J. Nucl. Phys.} 28 (1978) 822
(\emph{Yad. Fiz.} 28 (1978) 1597).

\bibitem{Iancu:2002tr} 
  E.~Iancu, K.~Itakura and L.~McLerran,
\emph{Geometric scaling above the saturation scale},
  \emph{Nucl.\ Phys.\ A} {\bf 708}, 327 (2002).
  
\bibitem{Caola:2008xr} 
  F.~Caola and S.~Forte,
\emph{Geometric Scaling from DGLAP evolution},
   \emph{Phys.\ Rev.\ Lett.}\  {\bf 101}, 022001 (2008).




  



\bibitem {Bartels:2002cj}J.~Bartels, K.~J.~Golec-Biernat and H.~Kowalski,
\emph{A modification of the saturation model: DGLAP evolution},
\emph{Phys.\ Rev.\ D} \textbf{66}, 014001 (2002) [arXiv:hep-ph/0203258].

  
\bibitem {Kowalski:2010ue}H.~Kowalski, L.~N.~Lipatov, D.~A.~Ross and G.~Watt,
\emph{Using HERA Data to Determine the Infrared Behaviour of the BFKL
Amplitude}, \emph{Eur.\ Phys.\ J.\ C }\textbf{70}, 983 (2010) [arXiv:1005.0355
[hep-ph]].

\bibitem{Beuf:2008mb}
  G.~Beuf,
 \emph{An Alternative scaling solution for high-energy QCD saturation with running coupling},
  arXiv:0803.2167 [hep-ph].
  \bibitem{DiffScal}
  E.~Iancu, A.~H.~Mueller and S.~Munier,
  \emph{Universal behavior of QCD amplitudes at high energy from general tools of statistical physics},
  \emph{Phys.\ Lett.\ B} {\bf 606} (2005) 342
  [hep-ph/0410018];\newline
  Y.~Hatta, E.~Iancu, C.~Marquet, G.~Soyez and D.~N.~Triantafyllopoulos,
\emph{Diffusive scaling and the high-energy limit of deep inelastic scattering in QCD at large} $N_c$,
 \emph{ Nucl.\ Phys.\ A} {\bf 773} (2006) 95
  [hep-ph/0601150].
  


\bibitem {Gelis:2006bs}F.~Gelis, R.~B.~Peschanski, G.~Soyez and L.~Schoeffel,
\emph{Systematics of geometric scaling}, \emph{Phys.\ Lett.\ B} \textbf{647}
(2007) 376 [hep-ph/0610435];\newline G.~Beuf, R.~Peschanski, C.~Royon and
D.~Salek, \emph{Systematic Analysis of Scaling Properties in Deep Inelastic
Scattering}, \emph{Phys.\ Rev.\ D} \textbf{78} (2008) 074004 [arXiv:0803.2186
[hep-ph]].




\bibitem {Beuf:2008bb}G.~Beuf, C.~Royon and D.~Salek, \emph{Geometric Scaling
of $F_{2}$ and $F_{2}^{c}$ in data and QCD Parametrisations}, [arXiv:0810.5082
[hep-ph]].




\bibitem {Royon:2010tz}C.~Royon and R.~Peschanski, \emph{Studies of scaling
properties in deep inelastic scattering}, \emph{PoS DIS} \textbf{2010} (2010)
282 [arXiv:1008.0261 [hep-ph]].


  
 \bibitem {HERAcombined}F.~D.~Aaron \textit{et al.} [H1 and ZEUS
Collaboration],
\emph{Combined Measurement and QCD Analysis of the Inclusive ep Scattering Cross
Sections at HERA},
\emph{JHEP} \textbf{1001} (2010) 109
[arXiv:0911.0884 [hep-ex]].

\bibitem{Praszalowicz:2013uu} 
  M.~Praszalowicz,
\emph{Violation of Geometrical Scaling in pp Collisions at NA61/SHINE},
  arXiv:1301.4647 [hep-ph].


\bibitem {GSinpp}L.~McLerran, M.~Praszalowicz, \emph{Saturation and
Scaling of Multiplicity, Mean $p_{\mathrm{T}}$, $p_{\mathrm{T}}$ Distributions
from 200 GeV $< \sqrt{s}$ 7 TeV}, \emph{Acta Phys.\ Pol.\ B} \textbf{41}
(2010) 1917 [arXiv:1006.4293 [hep-ph]] and
\newline\emph{Saturation and Scaling of
Multiplicity, Mean $p_{\mathrm{T}}$, $p_{\mathrm{T}}$ Distributions from 200
GeV $< \sqrt{s}$ 7 TeV -- Addendum}, \emph{Acta Phys.\ Polon.\ B} \textbf{42}
(2011) 99 [arXiv:1011.3403 [hep-ph]];\\
M.~Praszalowicz, \emph{Improved Geometrical
Scaling at the LHC}, \emph{Phys.\ Rev.\ Lett.}\ \textbf{106} (2011) 142002
[arXiv:1101.0585 [hep-ph]].



\end{thebibliography}
\end{document}